%% file: main.tex
\documentclass[gmd, manuscript]{copernicus}

\begin{document}

\nolinenumbers

\input{sections/_numbers}

\title{FESOM2.8 in single precision: climate fidelity over six decades and
performance on CPUs and GPUs}

\Author[1][suvarchal.cheedela@awi.de]{Suvarchal Kumar}{Cheedela}
\Author[1][nikolay.koldunov@awi.de]{Nikolay V.}{Koldunov}
\Author[1]{Sergey}{Danilov}
\Author[1]{Jan}{Streffing}
\Author[1]{Dmitry}{Sidorenko}
\Author[1,2]{Thomas}{Jung}

\affil[1]{Alfred Wegener Institute, Helmholtz Centre for Polar and Marine
Research, Bremerhaven, Germany}
\affil[2]{Institute for Environmental Physics, University of Bremen,
Bremen, Germany}

\runningtitle{FESOM2.8 in single precision}
\runningauthor{Cheedela et al.}

\correspondence{Suvarchal Kumar Cheedela (suvarchal.cheedela@awi.de)}

\received{}
\pubdiscuss{}
\revised{}
\accepted{}
\published{}

\firstpage{1}
\maketitle

\begin{abstract}
Halving the floating-point precision of an ocean model can make it
substantially faster, but whether the rounding errors stay negligible over
climate time scales has rarely been tested. We implement single precision in
the global ocean--sea-ice model FESOM2 and compare 60-year hindcasts with
double-precision controls. A first implementation lost salt steadily,
accumulated a heat-content deficit of \ohcGapSpDp\,ZJ, and shifted major
ocean fronts. Controlled experiments trace the drift to the tracer update:
small increments added to a large salinity value are rounded away, and the
increments themselves are biased. Accumulating tracer tendencies and layer
thicknesses in 64~bits, and carrying the rounding error of each tracer update
into the next, removes the drift for at most \stackCostFortranCpuMaxPct\,\%
additional runtime. Single- and
double-precision simulations then differ in heat content and mean state about
as much as two double-precision runs with slightly perturbed initial
conditions, sea ice differs by at most \finIceMaxPct\,\%, and the salinity
difference is four orders of magnitude smaller than before. In the coupled
model AWI-ESM3, single precision adds no detectable heat source or sink over
\awicmRunYears\ years.
Single precision speeds up the 60-year hindcasts by a factor of
\finProdSpeedup, and the model by up to \speedupFortranCpuMax\ on meshes of
up to 5\,km resolution; on GPUs, a C++/Kokkos version runs up to
\speedupPortGpuMax\ times faster and needs \gpuMemSavingMax\,\% less
memory. Whether reduced precision is safe for climate integrations is decided
by long-term budgets, not by short tests.
\end{abstract}

\input{sections/01_introduction}
\input{sections/02_model}

\input{sections/03_implementation}
\input{sections/04_fidelity}
\input{sections/05_performance}
\input{sections/06_discussion}
\input{sections/07_conclusions}

\input{sections/99_availability}

\appendix
\input{sections/A1_appendix}
\input{sections/A2_before}
\input{sections/A3_coupled}
\noappendix

\input{sections/99b_statements}

\bibliographystyle{copernicus}
\bibliography{references}

\end{document}

%% file: sections/_numbers.tex
\newcommand{\yardSstRmseObs}{0.619}  %
\newcommand{\yardSstRmsDiff}{0.165}  %
\newcommand{\yardSstRmsDiffAnom}{0.0172}  %
\newcommand{\yardSstRmsFloor}{0.0046}  %
\newcommand{\yardSstRatio}{0.27}  %
\newcommand{\yardSstRatioAnom}{0.028}  %
\newcommand{\yardSstOverFloor}{36}  %
\newcommand{\yardSstOverFloorAnom}{3.7}  %
\newcommand{\yardSstObs}{HadISST1}  %
\newcommand{\yardSstRmsFloorPert}{0.0040}  %
\newcommand{\yardSssRmseObs}{0.412}  %
\newcommand{\yardSssRmsDiff}{0.0497}  %
\newcommand{\yardSssRmsDiffAnom}{0.00883}  %
\newcommand{\yardSssRmsFloor}{0.0017}  %
\newcommand{\yardSssRatio}{0.12}  %
\newcommand{\yardSssRatioAnom}{0.021}  %
\newcommand{\yardSssOverFloor}{30}  %
\newcommand{\yardSssOverFloorAnom}{5.3}  %
\newcommand{\yardSssObs}{WOA09}  %
\newcommand{\yardSssRmsFloorPert}{0.0015}  %
\newcommand{\yardIceRatioMin}{0.029}  %
\newcommand{\yardIceRatioMax}{0.17}  %
\newcommand{\yardIceRatioAnomMax}{0.012}  %
\newcommand{\yardIceOverFloorMin}{7.5}
\newcommand{\yardIceOverFloorMax}{28}
\newcommand{\yardIceOverFloorAnomMax}{2.4}
\newcommand{\saltDriftSpDp}{6.8}  %
\newcommand{\saltDriftSpDpAnom}{0.21}  %
\newcommand{\saltDriftFloor}{-0.000005}  %
\newcommand{\saltDecompShift}{0.0013}  %
\newcommand{\saltTrendSpDpPerYear}{120}  %
\newcommand{\saltLossPerYearSci}{1.2\times10^{-4}}  %
\newcommand{\saltTrendSpDpAnomPerYear}{3.9}  %
\newcommand{\saltTrendFloorPerYear}{0.0002}  %
\newcommand{\saltAnomGain}{32}  %
\newcommand{\dpOwnSaltTrend}{0.0039}  %
\newcommand{\saltDriftOldPair}{5.5}  %
\newcommand{\fluxMarginOrders}{4.8}  %
\newcommand{\tempDriftSpDp}{24}  %
\newcommand{\tempDriftSpDpAnom}{1.6}  %
\newcommand{\tempDriftFloor}{-0.2}
\newcommand{\dpOwnTempTrend}{77}  %
\newcommand{\ohcGapSpDp}{129}  %
\newcommand{\ohcGapSpDpAnom}{8.6}  %
\newcommand{\ohcGapFloor}{-1}
\newcommand{\ohcAnomRemovedPct}{93}  %
\newcommand{\ohcSurfaceSpDp}{123}  %
\newcommand{\ohcInternalSpDp}{5.3}  %
\newcommand{\ohcInternalShareSpDp}{4}  %
\newcommand{\ohcSurfaceSpDpAnom}{4.7}  %
\newcommand{\ohcInternalSpDpAnom}{13.3}  %
\newcommand{\ohcInternalFloorMax}{0.004}  %
\newcommand{\heatSinkSpWm}{0.008}  %
\newcommand{\heatSinkSpAnomWm}{0.02}  %
\newcommand{\dpBudgetClosureWm}{0.002}  %
\newcommand{\sxSinkWm}{0.023}  %
\newcommand{\sxSaltUpsu}{2.7}  %
\newcommand{\sxKahanSourceWm}{0.0017}  %
\newcommand{\sxKahanSaltLossUpsu}{0.43}  %
\newcommand{\sxKahanCostPct}{0.9}  %
\newcommand{\sxFctAbsReductionPct}{41}  %
\newcommand{\sxFctAnomChangePct}{1}  %
\newcommand{\sxAbsKahanSaltUpsu}{54}  %
\newcommand{\emergeSaltYear}{1}  %
\newcommand{\emergeOhcAnomYear}{9}  %
\newcommand{\emergeBudgetYear}{1}  %
\newcommand{\emergeHeatSinkYear}{1}  %
\newcommand{\IceVolNhSpDpPct}{1.0}  %
\newcommand{\IceVolNhSpDpMaxPct}{1.5}  %
\newcommand{\IceVolNhAnomMaxPct}{0.22}
\newcommand{\IceVolNhFloorMaxPct}{0.05}
\newcommand{\IceVolShSpDpPct}{0.044}  %
\newcommand{\IceVolShSpDpMaxPct}{7.4}  %
\newcommand{\IceVolShAnomMaxPct}{0.19}
\newcommand{\IceVolShFloorMaxPct}{0.1}
\newcommand{\twinVanillaTwenty}{-1.87}  %
\newcommand{\twinLedgerTwenty}{1.85}  %
\newcommand{\twinAnomTwenty}{0.064}  %
\newcommand{\twinDpCtlMax}{0.0010}  %
\newcommand{\twinQuantTwo}{-0.14}  %
\newcommand{\twinAnomSuppression}{29}  %
\newcommand{\expRunSha}{27ed9f4f}  %
\newcommand{\expRanks}{512}  %
\newcommand{\expPertRanks}{864}  %
\newcommand{\prodSpeedupMedian}{1.41}  %
\newcommand{\prodSpeedupFinalYear}{1.43}
\newcommand{\prodSypdDp}{58.3}
\newcommand{\prodSypdSp}{82.5}
\newcommand{\prodWallDp}{1481}  %
\newcommand{\prodWallSp}{1047}
\newcommand{\anomCostPct}{0.4}  %
\newcommand{\expMergeCommit}{ea3bfebe}  %
\newcommand{\expPrHeadSha}{7c43e359}  %
\newcommand{\finSha}{50467fa0}  %
\newcommand{\finSaltDiffUpsu}{0.47}  %
\newcommand{\finSaltDiffAnomUpsu}{0.48}  %
\newcommand{\finSaltOverFloor}{14}  %
\newcommand{\finSaltFloorUpsu}{0.08}  %
\newcommand{\finSaltReduction}{14000}  %
\newcommand{\finSaltNonfluxPct}{0.4}  %
\newcommand{\finSaltRelChange}{1.5\times10^{-8}}  %
\newcommand{\finTempDiffMk}{0.009}  %
\newcommand{\finTempDiffAnomMk}{0.2}
\newcommand{\finOhcDiff}{0.05}  %
\newcommand{\finOhcDiffAnom}{1.0}
\newcommand{\finOhcMaxAbs}{1.5}  %
\newcommand{\finOhcFloorMaxAbs}{1.0}  %
\newcommand{\finIceVolNhPct}{0.03}  %
\newcommand{\finIceMaxPct}{0.2}  %
\newcommand{\finIceFloorMaxPct}{0.07}  %
\newcommand{\finSstRms}{0.0046}  %
\newcommand{\finSstRmsFloor}{0.0040}  %
\newcommand{\finSstOverFloor}{1.1}  %
\newcommand{\finSstOverFloorAnom}{0.97}
\newcommand{\finSstRatioPct}{0.7}  %
\newcommand{\finSssRms}{0.0015}  %
\newcommand{\finSssRmsFloor}{0.0016}  %
\newcommand{\finSssOverFloor}{0.98}  %
\newcommand{\finSssOverFloorAnom}{1.0}
\newcommand{\finSssRatioPct}{0.4}  %
\newcommand{\finHeatInternalWm}{7.3\times10^{-6}}  %
\newcommand{\finHeatInternalMaxZJ}{0.006}  %
\newcommand{\finDpResidualMax}{1.8\times10^{-15}}  %
\newcommand{\finSpResidualMax}{1.8\times10^{-10}}  %
\newcommand{\finHeatJumpMin}{1.3\times10^{-9}}  %
\newcommand{\finHeatJumpMax}{1.8\times10^{-9}}
\newcommand{\finProdSpeedup}{1.35}  %
\newcommand{\finWallDp}{1662}  %
\newcommand{\finWallSp}{1227}
\newcommand{\finDpVsEarlierSstRms}{0.007}  %
\newcommand{\finSypdDp}{52}
\newcommand{\finSypdSp}{70}
\newcommand{\compSpeedup}{1.56}  %
\newcommand{\compRepSpreadMaxPct}{0.09}  %
\newcommand{\compOceanSpeedup}{1.76}  %
\newcommand{\compIceSpeedup}{1.14}  %
\newcommand{\compTracerSpeedup}{1.85}
\newcommand{\speedupFortranCpuMin}{1.02}
\newcommand{\speedupFortranCpuMax}{1.66}
\newcommand{\speedupPortCpuMin}{1.12}
\newcommand{\speedupPortCpuMax}{1.50}
\newcommand{\speedupPortGpuMin}{1.05}
\newcommand{\speedupPortGpuMax}{1.27}
\newcommand{\ladderFortranCoreOneNode}{1.56}  %
\newcommand{\ladderFortranCoreHalfNode}{1.66}  %
\newcommand{\ladderFortranCoreMax}{1.02}  %
\newcommand{\ladderFortranCoreMaxRanks}{4096}  %
\newcommand{\ladderPortCpuCoreHalfNode}{1.46}
\newcommand{\ladderPortCpuCoreMax}{1.17}
\newcommand{\ladderPortGpuCoreOneNode}{1.19}  %
\newcommand{\ladderPortGpuCoreMax}{1.05}  %
\newcommand{\gpuNodeOverCpuNodeDp}{2.6}  %
\newcommand{\gpuNodeOverCpuNodeSp}{2.0}  %
\newcommand{\speedupFortranEddyMin}{1.27}  %
\newcommand{\speedupFortranEddyMax}{1.64}
\newcommand{\speedupFortranEddyLowMin}{1.40}  %
\newcommand{\speedupFortranEddyLowMax}{1.64}
\newcommand{\speedupFortranEddyHighMin}{1.27}  %
\newcommand{\speedupFortranEddyHighMax}{1.40}
\newcommand{\speedupFortranEddyPreMin}{1.18}  %
\newcommand{\speedupFortranEddyPreMax}{1.37}
\newcommand{\stackCostFortranCpuMinPct}{-2}  %
\newcommand{\stackCostFortranCpuMaxPct}{8}
\newcommand{\stackCostPortCpuMinPct}{-0.7}  %
\newcommand{\stackCostPortCpuMaxPct}{5}
\newcommand{\stackCostPortGpuMinPct}{0.5}  %
\newcommand{\stackCostPortGpuMaxPct}{3}
\newcommand{\stackCostFortranEddyTopPct}{5}  %
\newcommand{\legSpreadMedianPct}{0.3}  %
\newcommand{\legSpreadMaxPctRest}{3}  %
\newcommand{\legSpreadMaxPct}{12}  %
\newcommand{\ladderCampaignDpDiffMinPct}{-22}  %
\newcommand{\ladderCampaignDpDiffMaxPct}{30}
\newcommand{\speedupPortCpuEddyMin}{1.32}
\newcommand{\speedupPortCpuEddyMax}{1.50}
\newcommand{\sypdFCoreBestNodes}{16}  %
\newcommand{\sypdFCoreBestDp}{140}  %
\newcommand{\sypdFCoreBestSp}{150}  %
\newcommand{\sypdFCoreBestPre}{no}  %
\newcommand{\sypdFFarcBestNodes}{32}  %
\newcommand{\sypdFFarcBestDp}{59}  %
\newcommand{\sypdFFarcBestSp}{66}  %
\newcommand{\sypdFFarcBestPre}{no}  %
\newcommand{\sypdFDarsBestNodes}{256}  %
\newcommand{\sypdFDarsBestDp}{18}  %
\newcommand{\sypdFDarsBestSp}{22}  %
\newcommand{\sypdFDarsBestPre}{yes}  %
\newcommand{\sypdFNgBestNodes}{512}  %
\newcommand{\sypdFNgBestDp}{8.8}  %
\newcommand{\sypdFNgBestSp}{10}  %
\newcommand{\sypdFNgBestPre}{yes}  %
\newcommand{\sypdGCoreBestNodes}{2}  %
\newcommand{\sypdGCoreBestDp}{93}  %
\newcommand{\sypdGCoreBestSp}{110}  %
\newcommand{\sypdGCoreBestPre}{no}  %
\newcommand{\sypdGFarcBestNodes}{16}  %
\newcommand{\sypdGFarcBestDp}{41}  %
\newcommand{\sypdGFarcBestSp}{44}  %
\newcommand{\sypdGFarcBestPre}{no}  %
\newcommand{\sypdGDarsBestNodes}{32}  %
\newcommand{\sypdGDarsBestDp}{7.3}  %
\newcommand{\sypdGDarsBestSp}{8.4}  %
\newcommand{\sypdGDarsBestPre}{no}  %
\newcommand{\sypdGNgBestNodes}{32}  %
\newcommand{\sypdGNgBestDp}{4.2}  %
\newcommand{\sypdGNgBestSp}{4.8}  %
\newcommand{\sypdGNgBestPre}{no}  %
\newcommand{\partRecipeCostPct}{11}  %
\newcommand{\partRecipeCostSpPct}{12}
\newcommand{\partRecipeSpeedupDiag}{1.17}  %
\newcommand{\partRecipeSpeedupShipped}{1.18}  %
\newcommand{\partRecipeMatchPct}{2}  %
\newcommand{\gpuMemSavingMin}{40}  %
\newcommand{\gpuMemSavingMax}{43}
\newcommand{\gpuMemNgDevices}{16}  %
\newcommand{\gpuMemNgNodes}{4}  %
\newcommand{\gpuMemNgDpGiB}{35.9}  %
\newcommand{\gpuMemNgSpGiB}{21.6}  %
\newcommand{\portCpuSlowerMin}{5.8}  %
\newcommand{\portCpuSlowerMax}{23}
\newcommand{\gpuCoreOneNodeDpStep}{0.0679}  %
\newcommand{\cpuCoreHalfNodeDpStep}{0.381}  %
\newcommand{\faithSstRatio}{0.76}  %
\newcommand{\faithSstRatioErr}{0.08}
\newcommand{\faithSstFOverEnv}{1.6}  %
\newcommand{\faithSstPOverEnv}{2.4}
\newcommand{\faithSstF}{1.4\times10^{-3}}  %
\newcommand{\faithSstP}{1.1\times10^{-3}}  %
\newcommand{\faithSstBandRatio}{2.6}  %
\newcommand{\faithSstTropRatio}{0.54}  %
\newcommand{\faithSstAnomRatio}{0.35}  %
\newcommand{\faithSstAnomBandRatio}{1.2}
\newcommand{\faithSstCudaOverSerial}{1.06}  %
\newcommand{\faithSstCudaOverNoise}{5.1}  %
\newcommand{\faithTempRatio}{0.65}  %
\newcommand{\faithTempRatioErr}{0.06}
\newcommand{\faithTempFOverEnv}{1.5}  %
\newcommand{\faithTempPOverEnv}{2.4}
\newcommand{\faithTempF}{2.8\times10^{-3}}  %
\newcommand{\faithTempP}{1.8\times10^{-3}}  %
\newcommand{\faithTempBandRatio}{2.1}  %
\newcommand{\faithTempTropRatio}{0.52}  %
\newcommand{\faithTempAnomRatio}{0.32}  %
\newcommand{\faithTempAnomBandRatio}{1.1}
\newcommand{\faithTempCudaOverSerial}{1.12}  %
\newcommand{\faithTempCudaOverNoise}{4.9}  %
\newcommand{\faithSaltRatio}{0.83}  %
\newcommand{\faithSaltRatioErr}{0.09}
\newcommand{\faithSaltFOverEnv}{1.5}  %
\newcommand{\faithSaltPOverEnv}{2.6}
\newcommand{\faithSaltF}{1.7\times10^{-4}}  %
\newcommand{\faithSaltP}{1.4\times10^{-4}}  %
\newcommand{\faithSaltBandRatio}{1.9}  %
\newcommand{\faithSaltTropRatio}{0.62}  %
\newcommand{\faithSaltAnomRatio}{0.38}  %
\newcommand{\faithSaltAnomBandRatio}{1.1}
\newcommand{\faithSaltCudaOverSerial}{1.10}  %
\newcommand{\faithSaltCudaOverNoise}{5.2}  %
\newcommand{\faithSshRatio}{1.2}  %
\newcommand{\faithSshRatioErr}{0.05}
\newcommand{\faithSshFOverEnv}{2.7}  %
\newcommand{\faithSshPOverEnv}{7.0}
\newcommand{\faithSshF}{1.9\times10^{-3}}  %
\newcommand{\faithSshP}{2.3\times10^{-3}}  %
\newcommand{\faithSshBandRatio}{2.1}  %
\newcommand{\faithSshTropRatio}{0.54}  %
\newcommand{\faithSshAnomRatio}{0.33}  %
\newcommand{\faithSshAnomBandRatio}{0.91}
\newcommand{\faithSshCudaOverSerial}{1.00}  %
\newcommand{\faithSshCudaOverNoise}{12}  %
\newcommand{\faithAiceRatio}{1.9}  %
\newcommand{\faithAiceRatioErr}{0.06}
\newcommand{\faithAiceFOverEnv}{2.4}  %
\newcommand{\faithAicePOverEnv}{4.1}
\newcommand{\faithAiceF}{2.6\times10^{-3}}  %
\newcommand{\faithAiceP}{5.1\times10^{-3}}  %
\newcommand{\faithAiceBandRatio}{2.1}  %
\newcommand{\faithAiceAnomRatio}{1.4}  %
\newcommand{\faithAiceAnomBandRatio}{1.4}
\newcommand{\faithAiceCudaOverSerial}{1.02}  %
\newcommand{\faithAiceCudaOverNoise}{4.8}  %
\newcommand{\faithThreeYearMin}{0.52}  %
\newcommand{\faithThreeYearMax}{0.83}
\newcommand{\faithDpGapYearThree}{1.0\times10^{-2}}  %
\newcommand{\faithFortranSstYearThree}{5.3\times10^{-3}}
\newcommand{\faithPortSstYearThree}{3.4\times10^{-3}}
\newcommand{\faithCudaOverSerialMaxDevPct}{12}  %
\newcommand{\faithCudaOverNoiseMin}{4.8}
\newcommand{\footprintFiles}{51}
\newcommand{\footprintAdded}{1815}
\newcommand{\footprintRemoved}{657}
\newcommand{\footprintSrcFiles}{36}
\newcommand{\footprintSrcAdded}{1354}
\newcommand{\footprintSrcRemoved}{587}
\newcommand{\portBinSha}{dfab63c}  %
\newcommand{\oracleSha}{a62f180}  %
\newcommand{\ladderFortranSha}{50467fa0}  %
\newcommand{\portLadderSha}{010d423}  %
\newcommand{\meshCoreNodes}{126\,858}  %
\newcommand{\hindcastYears}{60}  %
\newcommand{\awicmRunYears}{60}  %
\newcommand{\awicmInternalAll}{-2.1\times10^{-4}}  %
\newcommand{\awicmInternalSpDp}{5.3\times10^{-7}}  %
\newcommand{\awicmInternalTwinDp}{1.2\times10^{-6}}  %
\newcommand{\awicmInternalTwinSp}{1.2\times10^{-6}}  %
\newcommand{\awicmResidSpMax}{2\times10^{-10}}  %
\newcommand{\awicmResidDpMax}{4\times10^{-15}}  %
\newcommand{\awicmHeatJump}{1.3\times10^{-9}}  %
\newcommand{\awicmOhcSpDp}{+24}  %
\newcommand{\awicmSstSpDp}{+0.11}  %
\newcommand{\awicmTasSpDp}{+0.22}  %
\newcommand{\awicmToaSpDp}{-0.20}  %
\newcommand{\awicmIceNhSpDp}{-22}  %
\newcommand{\awicmIceShSpDp}{+5}  %
\newcommand{\awicmOhcTwinDp}{+22}  %
\newcommand{\awicmSstTwinDp}{+0.05}  %
\newcommand{\awicmTasTwinDp}{+0.14}  %
\newcommand{\awicmToaTwinDp}{-0.15}  %
\newcommand{\awicmIceNhTwinDp}{-26}  %
\newcommand{\awicmIceShTwinDp}{+13}  %
\newcommand{\awicmOhcTwinSp}{-32}  %
\newcommand{\awicmSstTwinSp}{+0.01}  %
\newcommand{\awicmTasTwinSp}{+0.03}  %
\newcommand{\awicmToaTwinSp}{+0.10}  %
\newcommand{\awicmIceNhTwinSp}{+0.2}  %
\newcommand{\awicmIceShTwinSp}{-5}  %
\newcommand{\awicmOhcTwinMaxAbs}{54}  %
\newcommand{\awicmTracerSpeedup}{1.8}  %
\newcommand{\awicmIceSpeedup}{1.07}  %
\newcommand{\awicmModelSpeedup}{1.25}  %
\newcommand{\awicmWaitDp}{12}  %
\newcommand{\awicmWaitSp}{20}  %

%% file: sections/01_introduction.tex
\introduction
\label{sec:introduction}

Most ocean models compute in 64-bit floating-point arithmetic, although their discretization, forcing, and parameterization errors exceed the relative rounding error of 32-bit arithmetic, about $10^{-7}$, by orders of magnitude. The case for 32 bits is one of data movement. Kernels that perform few arithmetic operations per byte they load are limited by memory bandwidth rather than by peak arithmetic throughput \citep{williams2009roofline}, and ocean-model kernels are of this kind. Memory has lagged processor speed for three decades \citep{wulf1995memorywall}, and peak arithmetic throughput has outgrown DRAM bandwidth over the last two \citep{gholami2024memorywall}. Storing a field in 32 instead of 64 bits halves the bytes it moves and the memory it occupies, on CPUs and GPUs alike, and a vector instruction processes twice as many values. Numerical weather prediction has taken this step: the ECMWF Integrated Forecasting System runs its forecasts in single precision with no measurable loss of skill and about 40\,\% lower cost \citep{vana2017ifssp,lang2021moreaccuracy}. Reduced-precision emulation has identified the components that need more precision \citep{dueben2014inexact, dawson2017rpe, klower2020formats}, and single- and mixed-precision dynamical cores run in global atmospheric models \citep{nakano2018nicam, maynard2019endgame, chen2024grist}.

Ocean models have followed more slowly \citep{porter2025gpu}. Mixed-precision
versions of NEMO have been derived with emulation-based searches
\citep{tintoprims2019nemo,paronuzzi2025nemo}, and ECMWF now runs its NEMO
ocean--sea-ice forecasts in single precision while keeping double precision
for the ocean analysis \citep{keeley2024nemo4,polichtchouk2025cy50r1}. The
caution has a reason. The ocean integrates small fluxes over decades, and a
rounding bias that is invisible in a forecast can accumulate in the heat and
salt content. Published evaluations of reduced precision in ocean models cover
days to a few simulated years
\citep{tintoprims2019nemo,hafner2021veros,paronuzzi2025nemo}, but longer tests
show that such drifts occur: single-precision land-surface models
underestimate the slow warming of deep soil
\citep{harvey2016soil,dawson2018land}, and single-precision tracer advection in
a coarse-resolution ocean model shifted zonal-mean temperature by up to
1\,$^\circ$C over 300 years \citep{nakano2025som}. To our knowledge, the heat
and salt budgets of an ocean--sea-ice model run entirely in single precision
have not been evaluated over multiple decades, nor has the drift been traced to its source and removed.

Here we implement single working precision in FESOM2, an unstructured-mesh global ocean--sea-ice model \citep{danilov2017fesom2}, and ask whether it can be trusted for climate integrations and what it gains. By single precision we mean the model's default working precision; a small number of operations are kept in 64 bits (Sects.~\ref{sec:fp64-islands} and~\ref{sec:conservation}), so the implementation is mixed precision in the strict sense, with 32 bits as the default. Sixty-year hindcasts,
judged against double-precision runs that differ only in round-off or in tiny
initial perturbations, show that a straightforward implementation drifts in
salinity and heat content. We trace the drift to the way tracers are updated
and describe the changes now in FESOM2 that remove it, after which single and
double precision differ in the simulated climate not more than two
double-precision runs. Storing salinity relative to a reference value, as
early versions of the Modular Ocean Model did
\citep{pacanowski1999mom3}, reduced the drift but is no longer needed. We
then measure the speedup on four meshes from nominal $1^\circ$ to 5\,km
resolution, for the Fortran model on CPUs and for an independent C++/Kokkos
implementation on GPUs, and use the comparison of the two implementations as
a further test of numerical consistency. Finally, we repeat the budget test in
the coupled model AWI-ESM3.

%% file: sections/02_model.tex
\section{FESOM2}
\label{sec:model}

FESOM2 solves the hydrostatic, Boussinesq primitive equations on an
unstructured triangular mesh using a cell-vertex finite-volume discretization
\citep{danilov2017fesom2}. Temperature, salinity, pressure, and sea-surface
height are defined at mesh vertices; horizontal velocity is defined at
triangle centroids. We use the $z^\star$ form of the arbitrary
Lagrangian--Eulerian (ALE) vertical coordinate \citep{adcroft2004rescaled}.
The single-class dynamic--thermodynamic sea-ice model is based on FESIM
\citep{danilov2015fesim}, shares the ocean's scalar grid, and uses an
elastic--viscous--plastic (EVP) rheology. Previous studies describe the model
and its evaluation \citep{scholz2019fesom2,scholz2022fesom2} and computational
scalability \citep{koldunov2019fesom2scaling}. FESOM2 is the ocean component of
the AWI climate model \citep{streffing2022awi}; here it is run with prescribed
atmospheric forcing, except in Sect.~\ref{sec:coupled}.

The climate experiments (Sect.~\ref{sec:fidelity}) use the CORE2 mesh
(\meshCoreNodes\ surface vertices, 47 layers, nominal $1^\circ$ resolution)
and CORE-II interannual forcing for 1948--2007
\citep{large2009bulk,danabasoglu2014core2}.
The configuration uses K-profile parameterization (KPP) for vertical mixing
\citep{large1994kpp}, Gent--McWilliams eddy parameterization and Redi
isoneutral diffusion \citep{gent1990isopycnal,redi1982isopycnal},
backscatter-augmented viscosity \citep{juricke2020kinematic}, a 900\,s
timestep, and initial conditions from PHC3.0 \citep{steele2001phc}. The
timestep is half the 1800\,s used in production with this mesh, which makes
the test harder: the loss of small increments grows as the timestep shrinks
(Sect.~\ref{sec:unnoticed}). Performance measurements
(Sect.~\ref{sec:performance}) span four meshes from CORE2 to the
5-km-class NG5 mesh (Appendix~\ref{app:meshes}). These measurements and the
implementation comparison (Sect.~\ref{sec:port}) use JRA55-do forcing
\citep{tsujino2018jra55do}.

%% file: sections/03_implementation.tex
\section{Single precision in FESOM2}
\label{sec:implementation}

\subsection{A model-wide precision parameter}
\label{sec:wp-design}

FESOM2 declares its real variables with one kind parameter, \texttt{WP}.
A build option now sets it to 4~bytes and adjusts the compiler flags and the
MPI datatype of halo exchanges and reductions accordingly; the default
double-precision build is bit-identical to the code before the change. Single
precision thus covers the prognostic state, the mesh-derived operators, and
the ocean and sea-ice timestepping, with the exceptions described below. It
was first released in FESOM~2.8.0 and changed \footprintSrcFiles\ model-source
files, mostly literal kinds, interfaces, and input--output declarations that
assumed 64-bit reals. It is supported for standalone runs and for coupling to
OpenIFS through OASIS; other coupled configurations and the REcoM
biogeochemistry model, whose precision parameters have not been converted, are
rejected at build time. The model reports its working precision and source
revision at startup, and continuous integration builds both precisions.

\subsection{Operations kept in double precision}
\label{sec:fp64-islands}

A few operations stay in double precision because rounding them would bias
the result or cost accuracy for no gain in speed. Input and restart files
remain in double precision and are converted on reading. Time means for output are accumulated in 64~bits,
because summing thousands of 32-bit samples can bias slowly varying fields
such as salinity.

Global sums that close the surface budgets are accumulated and reduced in
64~bits. They add $10^6$--$10^9$ terms whose small residual results from large
cancelling contributions, and global sums are the operations most sensitive to
rounding and summation order in parallel climate codes
\citep{he2001accurate}. The sea-surface-height solver likewise accumulates its
conjugate-gradient scalars in 64~bits while keeping vectors and matrices in
single precision, as the Krylov solver of the ENDGame dynamical core does
\citep{maynard2019endgame}. Under the $z^\star$
coordinate, the solver's matrix changes each step by $10^{-7}$--$10^{-6}$ of
its entries, close to the 32-bit relative spacing of $1.2\times10^{-7}$; these
changes are accumulated in a 64-bit copy and rounded once per step. Single
precision also uses a symmetric Jacobi preconditioner by default, which
reduces the number of solver iterations by a third.

The CVMix library \citep{griffies2015cvmix} has fixed 64-bit interfaces, so
its mixing schemes remain in double precision; this study uses FESOM2's own
KPP scheme, which follows the working precision. Finally, the $10^{-40}$
guards against division by zero in KPP are subnormal in 32~bits and were
flushed to zero by optimizing compilers, producing $0/0$ at the first step;
they are now $10^{-20}$.

\subsection{Conservative tracer and layer-thickness updates}
\label{sec:conservation}

FESOM2 updates tracers in flux form on an arbitrary Lagrangian--Eulerian
vertical grid. The tracer content $hT$ of a control volume with layer
thickness $h$ changes by the divergence of the advective and diffusive fluxes,
$h^{n+1}T^{n+1}=h^nT^n+\Delta t\,R^n$, and each flux through an edge or a
layer interface enters the two adjacent control volumes with opposite signs,
so that in exact arithmetic the global content changes only through the
boundary fluxes. Three roundings break this in single precision. First, a
product of flux and timestep that is rounded before it is added to the two
control volumes no longer cancels between them. Second, rounding the new
state changes the content it represents: at $S\approx35$\,psu, adjacent 32-bit
values are $3.8\times10^{-6}$\,psu apart, and a smaller increment is partly or
entirely lost. Third, rounding the new layer thickness at every step changes
the volume of each control volume, which violates volume conservation and,
through $hT$, tracer conservation.

A defect of the first kind was found early. The flux-corrected transport (FCT)
scheme obtained its low-order tendency as the difference of two states,
$h^{n+1}T_{\mathrm{LO}}-h^nT^n$, which loses small tendencies in single
precision. It now computes the tendency from the flux divergence, in single
precision since FESOM~2.8.0 and in both precisions since 2.8.1.
In five-day coupled experiments, the defect had produced a global heat
imbalance of $-0.3$ to $-0.5$\,W\,m$^{-2}$ in single precision, compared with
at most 0.01\,W\,m$^{-2}$ in double precision.

The current code avoids all three roundings. Tracer tendencies are accumulated
in 64~bits, so that fluxes cancel between control volumes; this includes the
advective and diffusive tendencies, the low-order FCT tendency, and the column
arrays of the implicit vertical advection and diffusion solvers. The new
state is computed in 64~bits, and its rounding error, weighted by the new
thickness, is kept in one additional 32-bit array per tracer, $r$, and added
at the next update:
\begin{equation}
C=h^nT^n+\Delta t\,R^n+r^n,\qquad
T^{n+1}=\mathrm{fl}_{32}\!\left(C/h^{n+1}\right),\qquad
r^{n+1}=C-h^{n+1}T^{n+1},
\label{eq:carry}
\end{equation}
where $C$ is evaluated in 64~bits and $\mathrm{fl}_{32}$ rounds to the
nearest 32-bit value. The content represented by the state and the remainder,
$h^{n+1}T^{n+1}+r^{n+1}$, thus equals the 64-bit content $C$ up to the
rounding of the small remainder itself. This is
compensated summation \citep{kahan1965} applied to the thickness-weighted
tracer content; the implicit vertical diffusion, which updates the state a
second time in each step, updates $r$ in the same way. Layer thicknesses are
stored in 64~bits. Their halo exchange sends each value as a 32-bit number and
its 32-bit remainder, so communication remains in single precision, and output
and restart files receive 32-bit copies. Whereas an automatic precision search
for NEMO kept tracer advection entirely in double precision
\citep{paronuzzi2025nemo}, FESOM2 keeps its tracer fields and operators in
single precision and holds only the tendencies and the thicknesses in 64~bits.

In double precision, these quantities are 64-bit anyway and $r$ is not used,
so results are unchanged. In single precision, the conservative updates
lengthen a timestep by at most \stackCostFortranCpuMaxPct\,\%
(Sect.~\ref{sec:scaling}). The remainder is not written to restart files, so
each restart discards less than one 32-bit spacing per control volume; in the
experiments of Sect.~\ref{sec:final}, this changes the global heat content by
at most $\finHeatJumpMax$ of its value per restart.

\subsection{Salinity representation}
\label{sec:anomaly}

A runtime option stores salinity as a deviation from a reference value,
$S'=S-35$\,psu. The Modular Ocean Model time-stepped $(S-35)/1000$ ``for
purposes of numerical accuracy'' \citep{pacanowski1999mom3}, and atmospheric
models use deviations from a reference state for similar reasons
\citep{ruedisuehli2014cosmo,skamarock2019wrf}. Over most of the ocean,
$|S'|\lesssim2$\,psu, so representable values are about thirty times closer
together than at 35\,psu. Without the conservative updates, this
representation was the main remedy for the drift of single precision
(Sect.~\ref{sec:before}); with them, it no longer changes the results
(Sect.~\ref{sec:final}).

The equation of state, the KPP buoyancy flux, the ice--ocean coupling, and
other operations that need absolute salinity add the reference value back,
and output remains in absolute salinity. Under the nonlinear free surface,
freshwater also dilutes the reference part, which enters the surface boundary
condition for $S'$ as a virtual salt flux with constant salinity
\citep{griffies2004mom4}; \citet{pacanowski1999mom3} describe the sign error
that arises without it. Configurations whose salinity operations have not
been adapted, such as ice-shelf cavities and icebergs, stop at setup when the
option is enabled. In double precision, the option changes results only
through round-off, which makes it a control for the climate experiments.

%% file: sections/04_fidelity.tex
\section{Climate fidelity over six decades}
\label{sec:fidelity}

\subsection{Experiments and double-precision controls}
\label{sec:exp-pair}

We judge single precision by whether its differences from double precision
exceed those between double-precision runs that differ only in round-off or
in tiny initial perturbations. Such comparisons are established in port
validation and consistency testing
\citep{rosinski1997rounding,baker2016popect,massonnet2020replicability,zeman2022ensemble}
and have been applied to single precision in a regional climate model
\citep{banderier2024cosmo}. All experiments are \hindcastYears-year hindcasts
for 1948--2007 in the CORE2 configuration of Sect.~\ref{sec:model}, on
\expRanks\ processes and with the same solver preconditioner in both
precisions.

The main experiments use the current code with the conservative updates of
Sect.~\ref{sec:conservation}: two single-precision runs, one with absolute
salinity and one with the salinity anomaly, and two double-precision runs, one
of them started from temperatures perturbed by $2\times10^{-4}$\,K. The
difference between the two double-precision runs is our measure of chaotic
divergence. The anomaly run is compared with the unperturbed
absolute-salinity double-precision run, as the two representations differ only
in round-off in double precision. The model records its global volume, heat,
and salt budgets---content, surface fluxes, and residual---every simulated
day.

An earlier set of hindcasts with the first implementation
(Sect.~\ref{sec:before}) has its own double-precision control: two runs that
differ only in the salinity representation, and hence only in round-off. The
chaotic divergence is similar in both sets: the 1988--2007 mean sea-surface
temperature (SST) of the double-precision runs differs by
\finSstRmsFloor\,$^\circ$C RMS in the current code and by
\yardSstRmsFloor\,$^\circ$C in the earlier one. Global integrals use fixed
reference layer thicknesses; the neglected surface-height correction is less
than $10^{-3}$ of the volume.

\subsection{Single precision with conservative updates}
\label{sec:final}

\begin{figure*}[tp]
\centering
\includegraphics[width=\textwidth]{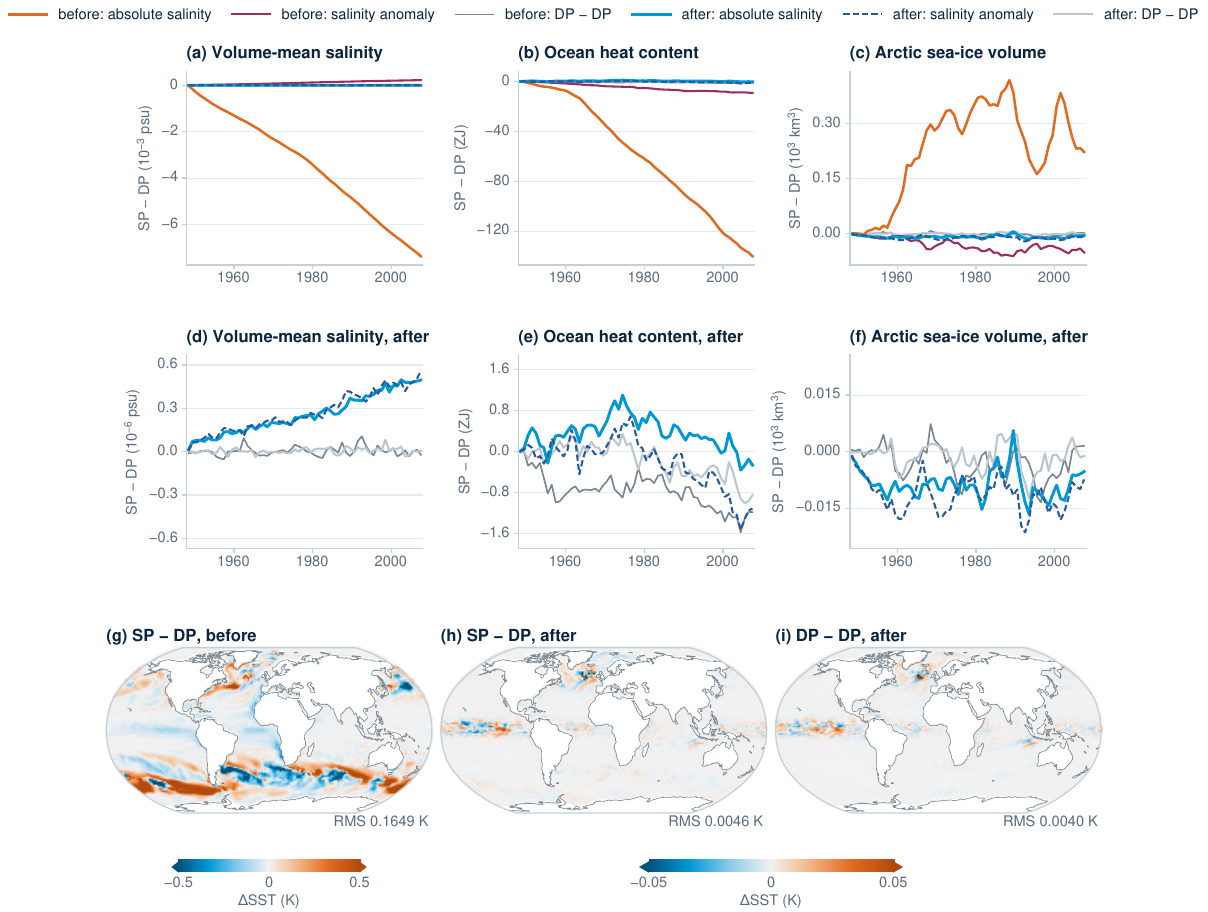}
\caption{Single minus double precision in the 60-year hindcasts before
(first implementation) and after the conservative updates of
Sect.~\ref{sec:conservation}. (a--c) Annual differences in volume-mean
salinity, ocean heat content, and Arctic sea-ice volume; gray lines are
differences between two double-precision runs: salinity anomaly minus
absolute salinity before (dark), perturbed minus unperturbed initial
temperatures after (light). (d--f) The runs after the updates and the
double-precision differences on a finer scale; note the unit in~(d).
(g--i) 1988--2007 mean SST difference before~(g) and after~(h) the updates,
and between the two double-precision runs after them~(i); (h) and~(i) share
a color scale ten times finer than that of~(g).
\label{fig:final}}
\end{figure*}

With conservative updates, single precision is indistinguishable from double
precision in heat content, temperature, and mean state (Fig.~\ref{fig:final}).
Over 60 years, the heat-content difference stays within \finOhcMaxAbs\,ZJ,
against \finOhcFloorMaxAbs\,ZJ between the double-precision runs, whereas the
first implementation lost \ohcGapSpDp\,ZJ. Sea ice differs slightly more than
between the double-precision runs: the annual volume and area of the Arctic
and Antarctic ice differ by at most \finIceMaxPct\,\%, against
\finIceFloorMaxPct\,\% (Fig.~\ref{fig:final}f). Separating
the heat-content difference into surface supply and internal residual
(Appendix~\ref{app:before-budgets}) confirms that no heat is created or
destroyed numerically: the internal residual is at most
$\finHeatInternalWm$\,W\,m$^{-2}$, compared with a sink of
\heatSinkSpAnomWm\,W\,m$^{-2}$ in the first implementation with the salinity
anomaly.

Salinity is the one quantity in which single precision remains
distinguishable. It gains salt at a nearly constant rate, reaching
$\finSaltDiffUpsu\times10^{-6}$\,psu in the last decade
(Fig.~\ref{fig:final}d), about \finSaltOverFloor\ times the RMS difference
between the double-precision runs but \finSaltReduction\ times less than the
first implementation lost.
The model's own budget shows that this salt enters through the surface: the
surface salt flux supplies all but \finSaltNonfluxPct\,\% of the difference in
salt content, and the volume and salt budgets close to within
$\finSpResidualMax$ of the content, compared with $\finDpResidualMax$ in double
precision. We have not identified why single precision takes up slightly more
salt at the surface.

The 1988--2007 mean state shows the same picture (Fig.~\ref{fig:final}g--i).
The single-minus-double-precision SST difference has the amplitude and pattern
of the difference between the double-precision runs, an RMS of
\finSstRms\,$^\circ$C against \finSstRmsFloor\,$^\circ$C, concentrated in the
tropical Pacific and the Gulf Stream. It amounts to \finSstRatioPct\,\% of the
model's error against HadISST1 \citep{rayner2003hadisst}; for surface
salinity, the corresponding figure is \finSssRatioPct\,\% of the error against
the World Ocean Atlas 2009 \citep{antonov2010woa09}. In the first
implementation, the differences were \yardSstOverFloor\ and
\yardSssOverFloor\ times the control.

Storing salinity as an anomaly no longer makes a measurable difference. The
salinity gain is the same, $\finSaltDiffAnomUpsu\times10^{-6}$ against
$\finSaltDiffUpsu\times10^{-6}$\,psu, and heat content and mean state again
differ about as much as the two double-precision runs.

\subsection{Single precision without conservative updates}
\label{sec:before}

The first implementation, without conservative updates, behaved differently
(Fig.~\ref{fig:final}a--c,~g; Appendix~\ref{app:before}). With absolute
salinity, it lost salt at $\saltLossPerYearSci$\,psu\,yr$^{-1}$ and
accumulated a heat-content deficit of \ohcGapSpDp\,ZJ, mostly through altered
surface fluxes; its mean state shifted along the major fronts, and Arctic sea
ice thickened. Controlled experiments ruled out the surface freshwater budget,
the global sums, the solver, and the domain decomposition, and traced the
drift to the tracer update itself. Keeping the state in 64~bits while
computing the increments in 32~bits produced a drift of similar size and
opposite sign, which showed that the increments, not only their addition to
the state, were biased. Storing salinity as an anomaly reduced the drift by a
factor of \saltAnomGain\ and the heat-content deficit by
\ohcAnomRemovedPct\,\%, but left a salinity gain and an internal heat sink of
\heatSinkSpAnomWm\,W\,m$^{-2}$. In nine-month experiments, compensated
summation of the update alone did not remove these residuals; the
conservative updates, which also compute the increments and the layer
thicknesses in 64~bits, do (Sect.~\ref{sec:final}).

%% file: sections/05_performance.tex
\section{Performance}
\label{sec:performance}

\subsection{CORE2 hindcasts}
\label{sec:anatomy}

In the 60-year hindcasts of Sect.~\ref{sec:final}, run on four nodes with full
output, single precision raises the throughput from \finSypdDp\ to
\finSypdSp\ simulated years per day, a speedup of \finProdSpeedup; the first
implementation, without conservative updates, reached \prodSpeedupMedian.
Storing salinity as an anomaly costs \anomCostPct\,\% of the runtime.

\begin{figure}[tp]
\includegraphics[width=8.3cm]{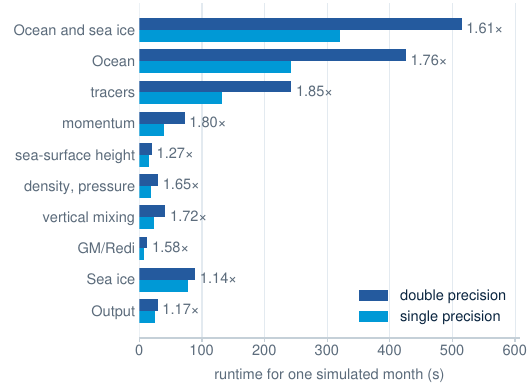}
\caption{Component runtimes for one simulated month of the CORE2 hindcast
configuration on one node (128 processes) in double and single precision,
each the faster of two runs, which differ by at most
\compRepSpreadMaxPct\,\%. Labels give the component speedups.
\label{fig:components}}
\end{figure}

The gain comes from the ocean (Fig.~\ref{fig:components}). In one simulated
month on one node, where single precision runs \compSpeedup\ times faster
overall, the ocean component gains a factor of \compOceanSpeedup\ and its
tracer calculations a factor of \compTracerSpeedup, whereas sea ice gains only
a factor of \compIceSpeedup. This is the pattern expected if the ocean timestep
is limited by memory traffic, which single precision halves, while the
sea-ice subcycles, with short loops and frequent synchronization, are limited
by latency.

\subsection{Scaling across meshes and hardware}
\label{sec:scaling}

To see how the gain depends on resolution and hardware, we measure strong
scaling on four meshes, from CORE2 to the 5-km NG5 mesh
(Appendix~\ref{app:meshes}), for the Fortran model on CPUs and for the
C++/Kokkos implementation (Sect.~\ref{sec:port}) on NVIDIA A100 GPUs of the
DKRZ Levante system. Each measurement runs double precision, single
precision, and single precision without conservative updates for 300
timesteps, twice each in mirrored order; repeated runs differ by a median of
\legSpreadMedianPct\,\% (Appendix~\ref{app:campaign}).

\begin{figure*}[tp]
\includegraphics[width=\textwidth]{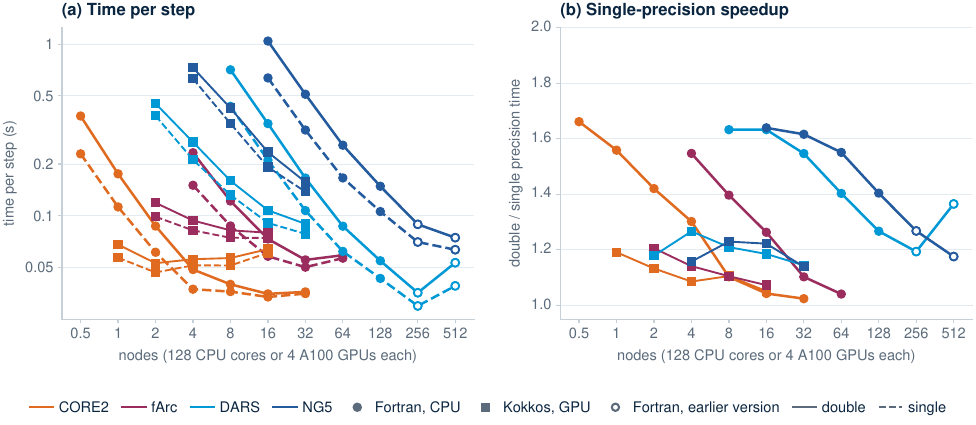}
\caption{Strong scaling in double (solid) and single (dashed) precision for
Fortran on CPUs and C++/Kokkos on GPUs. Each node has 128 CPU cores or four
A100 GPUs; the 64-process CORE2 point uses half a CPU node. Open symbols: an
earlier Fortran version without the conservative updates of
Sect.~\ref{sec:conservation}. (a) Time per step. (b) Single-precision speedup.
\label{fig:scaling}}
\end{figure*}

The speedup depends mainly on how much data each process holds
(Fig.~\ref{fig:scaling}). It is largest when the working set of each process
far exceeds the cache and shrinks as the working set fits into it. On CORE2,
the Fortran speedup falls from \ladderFortranCoreHalfNode\ on half a node to
\ladderFortranCoreMax\ on \ladderFortranCoreMaxRanks\ processes. The eddying DARS and NG5 meshes retain speedups of
\speedupFortranEddyLowMin--\speedupFortranEddyLowMax\ up to 64 nodes and
\speedupFortranEddyHighMin--\speedupFortranEddyHighMax\ at 128 nodes; at 256
and 512 nodes, measured with an earlier version without conservative updates,
the speedup is \speedupFortranEddyPreMin--\speedupFortranEddyPreMax. The
conservative updates cost at most \stackCostFortranCpuMaxPct\,\% where
computation dominates and less than 1\,\% at the largest process counts.

On GPUs, the Kokkos implementation gains less, a factor of
\speedupPortGpuMin--\speedupPortGpuMax, largest on DARS and smallest on
CORE2, where runs on more than two nodes are limited by latency. Single
precision also reduces device memory use by
\gpuMemSavingMin--\gpuMemSavingMax\,\%, so that NG5 on \gpuMemNgDevices\ GPUs
needs \gpuMemNgSpGiB\ instead of \gpuMemNgDpGiB\,GiB per device. On CORE2,
one GPU node runs \gpuNodeOverCpuNodeDp\ times faster than one CPU node with
the Fortran model in double precision and \gpuNodeOverCpuNodeSp\ times faster
in single precision, which gains more on CPUs.

\begin{figure*}[tp]
\includegraphics[width=\textwidth]{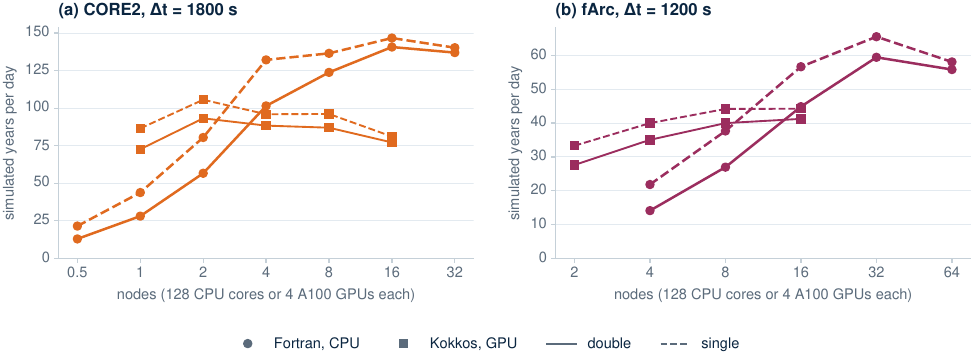}
\caption{Projected throughput in simulated years per day for CORE2 (a) and
fArc (b) in double (solid) and single (dashed) precision, for Fortran on CPUs
(circles) and Kokkos on GPUs (squares). CORE2 uses an 1800\,s timestep; fArc
throughput is projected from the 900\,s benchmark timestep to the 1200\,s
production timestep. \label{fig:sypdsmall}}
\end{figure*}

\begin{figure*}[tp]
\includegraphics[width=\textwidth]{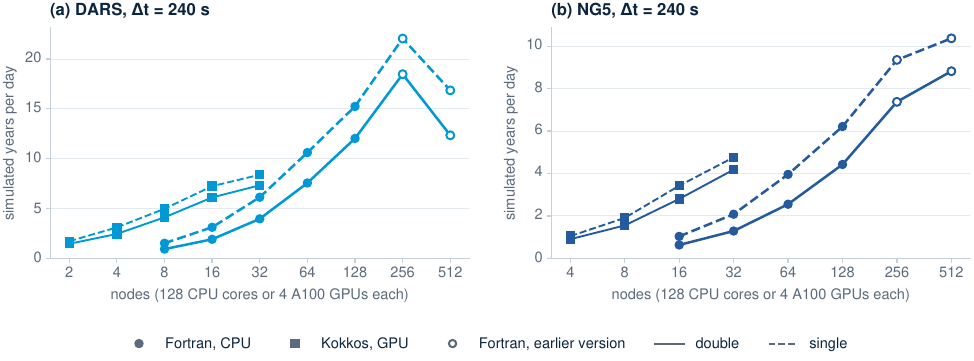}
\caption{As Fig.~\ref{fig:sypdsmall}, for DARS (a) and NG5 (b), projected from
benchmark timesteps of 120 and 180\,s to a 240\,s production timestep. Open
symbols: an earlier Fortran version (see Fig.~\ref{fig:scaling}).
\label{fig:sypdlarge}}
\end{figure*}

For production use, these gains translate into throughput
(Figs.~\ref{fig:sypdsmall} and~\ref{fig:sypdlarge}). At the fastest measured
configurations, single precision raises the projected Fortran throughput on
DARS from \sypdFDarsBestDp\ to \sypdFDarsBestSp\ simulated years per day on
\sypdFDarsBestNodes\ nodes and on NG5 from \sypdFNgBestDp\ to \sypdFNgBestSp\
on \sypdFNgBestNodes\ nodes, both measured without conservative updates;
Kokkos on GPUs reaches \sypdGDarsBestSp\ simulated years per day on DARS with
\sypdGDarsBestNodes\ nodes. These projections exclude output.
Configurations that use the CVMix library, which remains in double precision,
may gain less.

%% file: sections/06_discussion.tex
\section{Precision sensitivity of an independent implementation}
\label{sec:port}

FESOM2 is also being developed in C++/Kokkos \citep{trott2022kokkos} for GPUs,
verified kernel by kernel against a C reference that reproduces the long-term
statistics of the Fortran model \citep{koldunov2026llmport}, and in JAX for
differentiability, verified kernel by kernel against the Fortran model
\citep{koldunov2026fesomjax}. An independent implementation also allows a test that
double-precision verification cannot provide: whether reducing precision
changes two implementations of the same equations in the same way. We compare
one-year CORE2 integrations with JRA55-do forcing for 1958, using a Kokkos
version without conservative updates. Because the double-precision
trajectories of the two implementations diverge chaotically, each is compared
with itself: the precision effect is the relative $L_2$ distance
$\|X_{\mathrm{SP}}-X_{\mathrm{DP}}\|_2/\|X_{\mathrm{DP}}\|_2$ between its
single- and double-precision December fields, measured against a baseline
from five double-precision runs with initial perturbations of $10^{-6}$\,K
(Table~\ref{tab:faith}).

\begin{table*}[t]
\caption{Relative $L_2$ distances between single- and double-precision
December-1958 fields after one year. Each implementation uses its own
double-precision reference. The first numerical column gives the ratio of mean distances,
with uncertainty propagated from the standard deviations of four
single-precision members per implementation. The next two columns normalize
each implementation's mean distance by its double-precision variability
baseline. The last column gives the Kokkos-to-Fortran ratio with the salinity
anomaly enabled.\label{tab:faith}}
\begin{tabular}{lcccc}
\tophline
Field & Kokkos / Fortran & Fortran / baseline & Kokkos / baseline & Kokkos / Fortran, anomaly \\
\middlehline
SST & $\faithSstRatio \pm \faithSstRatioErr$ & \faithSstFOverEnv & \faithSstPOverEnv & \faithSstAnomRatio \\
temperature & $\faithTempRatio \pm \faithTempRatioErr$ & \faithTempFOverEnv & \faithTempPOverEnv & \faithTempAnomRatio \\
salinity & $\faithSaltRatio \pm \faithSaltRatioErr$ & \faithSaltFOverEnv & \faithSaltPOverEnv & \faithSaltAnomRatio \\
sea-surface height & $\faithSshRatio \pm \faithSshRatioErr$ & \faithSshFOverEnv & \faithSshPOverEnv & \faithSshAnomRatio \\
ice concentration & $\faithAiceRatio \pm \faithAiceRatioErr$ & \faithAiceFOverEnv & \faithAicePOverEnv & \faithAiceAnomRatio \\
\bottomhline
\end{tabular}
\belowtable{}
\end{table*}

For global temperature and salinity, the precision effect in Kokkos is
\faithTempRatio--\faithSaltRatio\ of that in Fortran, and on GPUs it agrees
with the CPU backend within \faithCudaOverSerialMaxDevPct\,\%. Reaching this
agreement exposed two defects that double-precision verification had missed:
an uninitialized Adams--Bashforth tracer history, and Redi/GM increments added
directly to salinity instead of to the accumulated tendency, which rounded
them at the spacing of $S\approx35$\,psu. Correcting the latter halved the
precision effect, as the advice to form the complete increment before adding
it to the state suggests \citep{croci2023heat,klower2022float16}. Larger
differences remain for sea-ice concentration and between 60 and
70$^\circ$S, where the Kokkos effect is \faithSstBandRatio\ times the Fortran
one for SST and \faithSaltBandRatio\ times for salinity. With the salinity
anomaly, these ratios fall to \faithSstAnomBandRatio\ and
\faithSaltAnomBandRatio, which points to rounding of absolute salinity under
Antarctic sea ice.

\section{Discussion}
\label{sec:discussion}

\subsection{When is reduced precision acceptable?}
\label{sec:acceptable}

Reduced precision changes a simulation in two ways. Rounding perturbations
grow chaotically, so single- and double-precision runs follow different
trajectories, which alone does not degrade a climate simulation. A biased
numerical source or sink, in contrast, accumulates. Global budgets tell the two
apart: salt changes only through the surface, so a salinity difference beyond
the controls that is not supplied by the surface flux is numerical, and the
heat budget separates a numerical heat source from the response of the
surface fluxes. We therefore propose two criteria. Conserved quantities should
show no numerical source or sink beyond the spread of the controls over the
length of the intended integration, and the simulated climate, including its
mean state and variability, should differ from double precision by much less
than the model differs from observations; individual trajectories may
diverge.

The first implementation failed the first criterion even where it met the
second: with the salinity anomaly, its mean-state differences were only
\yardSstRatioAnom\ and \yardSssRatioAnom\ of the errors in SST and surface
salinity, yet it gained salt and lost heat internally. With conservative
updates, both criteria are met over 60 years at nominal $1^\circ$ resolution
(Sect.~\ref{sec:final}). The remaining salinity difference exceeds the
controls but enters through the surface flux; whether a difference of this
size matters depends on the application, for example passive-tracer budgets
over centuries.

\subsection{Why short evaluations miss the drift, and how to avoid it}
\label{sec:unnoticed}

Several factors may explain why such drifts have not been reported for ocean
models. Some earlier models stored salinity relative to a reference value
(Sect.~\ref{sec:anomaly}). Mixed-precision designs keep the prognostic state in
double precision; the single-precision build of NEMO~4.2, for example,
declares temperature, salinity, velocities, and sea-surface height in double
precision \citep{nemo2024v422}. This removes the rounding of the update but,
as our experiments show, not necessarily the bias of increments computed from
absolute values (Sect.~\ref{sec:mechanism}). Evaluations have been short and
have compared fields or forecast skill rather than budgets. In budgets, the
drift is visible within the first model year (Sect.~\ref{sec:timeseries}), and
a fully single-precision NEMO configuration gained heat and salt relative to double precision within a month
\citep{tintoprims2019nemo}, but the climatic consequences develop over decades,
and the surface fluxes can hide a heat loss for years.

The underlying problem is general: small increments to a large background
value are poorly resolved in reduced precision, as documented for soil
temperatures in kelvin \citep{harvey2016soil,dawson2018land}, tracer advection
in an ocean model \citep{nakano2025som}, and 16-bit shallow-water models
\citep{klower2020formats}. Because the increment per step scales with the
timestep while the spacing of representable values does not, the error grows
as the timestep shrinks \citep{harvey2016soil,klower2022float16}, so
eddy-resolving configurations with timesteps of a few minutes would be more
exposed than our CORE2 hindcasts with their 900\,s timestep. A reference value reduces the
problem, as it did here by a factor of \saltAnomGain. Compensated summation of
the update \citep{kahan1965,moller1965,klower2022float16,lai2025mpasqdp}
matched a double-precision state over 300 years in an ocean advection scheme
\citep{nakano2025som}, and stochastic rounding
\citep{croci2022stochastic,paxton2022climate,kimpson2023climate} is unbiased in
expectation but needs hardware support or emulation. Compensated summation
alone, however, corrects only the rounding of the update and leaves biased
increments (Sect.~\ref{sec:mechanism}), and the Kokkos comparison shows that
where increments are accumulated matters as well. FESOM2 therefore combines
compensated updates with 64-bit tendencies and layer thicknesses, after which
the representation of salinity no longer matters. The same structure, a small
increment added each step to a large value, appears in dissolved inorganic
carbon and alkalinity and in the solver matrix under the $z^\star$ coordinate
(Sect.~\ref{sec:fp64-islands}).
How prognostic variables are stored and updated is part of reduced-precision
model design, not a compiler setting.

\subsection{Reduced precision as a verification tool}

Reduced precision amplifies errors that double precision masks, in particular
cancellations and the accumulation of small increments. Here it exposed an
uninitialized Adams--Bashforth history and Redi/GM increments added directly
to salinity in the C++/Kokkos implementation (Sect.~\ref{sec:port}), the
recovery of the FCT tendency as a difference of two states
(Sect.~\ref{sec:conservation}), and subnormal guards in KPP
(Sect.~\ref{sec:fp64-islands}). The first also affected double precision; the
others are specific to reduced precision. Comparing how independent
implementations respond to reduced precision, together with long-term budgets,
is therefore a sensitive check for models ported to new languages and
hardware.

\subsection{Single precision in a coupled model}
\label{sec:coupled}

Under prescribed atmospheric forcing, the surface fluxes respond to the ocean
only through the bulk formulae. In a coupled model, the atmosphere can
compensate a numerical heat sink in the ocean, or amplify it. We therefore
repeated the comparison in the coupled model AWI-ESM3, in which FESOM~2.8.3 on
the CORE2 mesh exchanges fluxes with the atmosphere model OpenIFS and the
vegetation model LPJ-GUESS (Appendix~\ref{app:coupled}). Four
\awicmRunYears-year pre-industrial control integrations differ only in the
precision of FESOM2 and in a perturbation of the initial ocean temperature by
$2\times10^{-4}$\,K. OpenIFS runs in single precision, its default, in all
four.

The budgets close as in the forced runs. The heat that the model's own fluxes
do not account for is $\awicmInternalAll$\,W\,m$^{-2}$ in all four runs and
does not depend on precision: single and double precision differ by
$\awicmInternalSpDp$\,W\,m$^{-2}$, the perturbed runs from their references by
$\awicmInternalTwinDp$ in double and $\awicmInternalTwinSp$\,W\,m$^{-2}$ in single
precision. In single
precision, the volume and salt residuals remain below $\awicmResidSpMax$ of
the content. The FCT defect (Sect.~\ref{sec:conservation}) would have shown up
here as a sink of $-0.3$ to $-0.5$\,W\,m$^{-2}$.

The coupled climate is a weaker test, because it is far noisier. With a freely
evolving atmosphere, the heat content of two runs from almost the same state
drifts apart by up to \awicmOhcTwinMaxAbs\,ZJ within decades, against
\finOhcFloorMaxAbs\,ZJ under prescribed forcing. Within this spread, single
and double precision are indistinguishable in heat content, global-mean
surface temperatures, the radiation balance at the top of the atmosphere, and
sea ice (Table~\ref{tab:coupled}).

Single precision speeds up the tracer calculations of FESOM2 by a factor of
\awicmTracerSpeedup\ but its sea ice only by \awicmIceSpeedup, the same
contrast as in the forced runs (Sect.~\ref{sec:anatomy}). With an unchanged layout, the coupled model
runs only \awicmModelSpeedup\ times faster, because FESOM2 then spends
\awicmWaitSp\ instead of \awicmWaitDp\,\% of its time waiting for the
atmosphere. We expect the larger gain from giving the ocean fewer processes at
the same throughput.
\subsection{Single precision as the default working precision}
\label{sec:default}

The results support making single precision the default working precision of an ocean model, with double precision reserved for the operations that need it. Three properties of current hardware make this the right default rather than an optional optimization.

First, ocean models are limited by memory traffic, not by arithmetic. Kernels that perform few floating-point operations per byte loaded are bounded by memory bandwidth long before peak arithmetic throughput \citep{williams2009roofline}, and ocean-model kernels are of this kind. Storing a field in 32 instead of 64 bits halves the bytes it moves and the memory it occupies, and doubles the fraction of the working set that fits in cache. This is the gain that the component timings show (Sect.~\ref{sec:anatomy}): the tracer calculations, which stream large arrays, gain a factor of \compTracerSpeedup, whereas sea ice, limited by latency, gains \compIceSpeedup.

Second, where arithmetic does matter, 32-bit arithmetic is at least twice as fast. A vector register holds twice as many 32-bit values, so a vectorized loop on the AMD EPYC processors used here processes eight instead of four values per instruction. On the A100 GPUs used here, the ratio of 32- to 64-bit throughput is two \citep{nvidia2020a100}; on NVIDIA's consumer and workstation GPUs of the Ada generation it is 64 \citep{nvidia2023ada}; and on NVIDIA's Blackwell Ultra 64-bit arithmetic was cut outright, to about 1.2\,TFLOP\,s$^{-1}$ per GPU against 37 on its predecessor, while low-precision matrix throughput grew; the announced Rubin generation restores 33\,TFLOP\,s$^{-1}$, at a 32- to 64-bit ratio of about four \citep{nvidia2026hgx}. On Blackwell Ultra, single precision is not a speedup but the condition for running at all.

Third, these gains address different bottlenecks and do not multiply. A kernel that is bandwidth-bound before conversion remains bandwidth-bound after it, so the speedup of a memory-bound model is close to two at best, and smaller where latency, integer index arrays, retained 64-bit arrays, or synchronization dominate. That is what we measure: \speedupFortranCpuMin--\speedupFortranCpuMax\ on CPUs and \speedupPortGpuMin--\speedupPortGpuMax\ on GPUs. The matrix units of recent processors, such as GPU tensor cores and the Arm and x86 matrix extensions, offer far more than twofold throughput for reduced precision, but only for dense matrix products, which the finite-volume loops of an ocean model do not contain, and partly in formats such as TF32 that carry fewer significand bits than IEEE single precision. They matter for solvers and data-driven components (Sect.~\ref{sec:remains}), not for the present implementation.

The trend strengthens the argument. Memory has lagged processor speed for three decades \citep{wulf1995memorywall}, and peak arithmetic throughput has outgrown DRAM bandwidth over the last two \citep{gholami2024memorywall}; between the two GPU generations in use at DKRZ, 64-bit throughput grew 3.5-fold while memory bandwidth grew by a factor of 1.6--2.2 \citep{nvidia2020a100, nvidia2022h100}. For a model limited by bytes rather than operations, the number of bytes per value is the one lever that keeps pace with the hardware, and reduced precision makes a larger share of a machine's capability usable for the same allocation.

Single precision in this sense is mixed precision. The working precision sets the default; the operations that keep 64 bits (Sects.~\ref{sec:fp64-islands} and~\ref{sec:conservation}) are the exceptions, chosen because rounding them creates a bias, and the conservative updates show how few they need to be. Mixed-precision ocean models have so far been derived from 64-bit code by a search over what can be demoted \citep{tintoprims2019nemo, paronuzzi2025nemo}, and the single-precision build of NEMO keeps the prognostic state in 64 bits \citep{nemo2024v422}. Holding the state in 32 bits as well captures the memory gain for the whole model and makes the cost of every exception explicit. The recommendation is conditional on validation: the budget diagnostics of Sect.~\ref{sec:acceptable} decide whether a configuration, resolution, or tracer needs an exception, and the exceptions remain part of the model's design, documented and tested, not left to the compiler.

\subsection{Outlook}
\label{sec:remains}

The speedups found here are of the size reported for NEMO, with about 30\,\% less time per step in mixed precision \citep{paronuzzi2025nemo} and up to 1.8 in single precision, diminishing near the scaling limit \citep{keeley2024nemo4}, and for Veros on GPUs \citep{hafner2021veros}. At kilometre scale, the smaller memory footprint on GPUs may matter as much as the speedup, because it leaves room for larger meshes, ensembles, or additional Earth-system components.

Separating the tracer update into tendency accumulation, layer geometry, and the conservative state update (Sect.~\ref{sec:conservation}) has made explicit where in the data flow precision matters, and the model-wide kind parameter lets each part be assigned its own. This opens the parts of the model that tolerate less than the working precision. The first candidate is the preconditioner of the sea-surface-height solver: its errors are corrected by the outer iteration as long as residuals and convergence tests are evaluated accurately, and half-precision preconditioning has been demonstrated for geophysical elliptic solvers \citep{ackmann2022solvers}. For the conjugate-gradient solver of FESOM2 this requires a preconditioner that remains symmetric positive definite, or a flexible variant of the iteration. The second is time integration: mixed-precision Runge--Kutta methods evaluate the costly stages, or all but the few that set the accuracy, in lower precision and retain the order and stability of the scheme \citep{grant2022perturbedrk, croci2022mixedrk}. Both must be judged by the budget diagnostics of this paper, and the second meets the small-increment problem head-on, because substeps shrink the increments further (Sect.~\ref{sec:unnoticed}). Beyond their own gains, these steps are what allow physics-based models to use processors whose reduced-precision units grow while their 64-bit units do not (Sect.~\ref{sec:default}), and to remain computationally competitive with data-driven climate models, which run in 16- or 32-bit precision on hardware built for them.

Reduced precision also changes the balance between local computation and communication. For a fixed problem size, faster local execution brings the model into the latency-dominated regime at a smaller process count: single precision reduces the volume of the halo exchanges but not their number, their start-up latency, or the number of synchronizations, which is why its speedup falls toward one at the largest process counts (Fig.~\ref{fig:scaling}b) and why sea ice gains least. Realizing the full benefit therefore calls for aggregating compatible halo exchanges into fewer messages, overlapping them with independent computation, and reducing the synchronizations in the iterative solver. An earlier scaling limit for a single integration does not limit throughput, however. Once additional processes no longer shorten an integration, the same resources run more concurrent ensemble members, each at an efficient process count. ECMWF likewise reinvested the savings of single precision in its ensemble, there in vertical resolution \citep{lang2021moreaccuracy}. Climate applications that need ensembles to characterize internal variability and the probabilities of extremes gain in simulated years per node-hour even where communication prevents a single run from going faster.

The conservative updates have been evaluated over 60 years at nominal $1^\circ$ resolution only. Eddy-rich and kilometre-scale configurations combine shorter timesteps with stronger intrinsic variability, which can mask precision errors \citep{penduff2011intrinsic, leroux2018amoc, nakano2025som}. Century-scale conservation and passive tracers remain untested. The biogeochemistry model REcoM still declares its own 64-bit working precision, so FESOM2 builds that include it, such as the carbon-cycle configuration of AWI-ESM3, stay in double precision. Making REcoM follow the working precision is the next step; its dissolved inorganic carbon and alkalinity will need the same conservative updates as salinity (Sect.~\ref{sec:mechanism}) and their own budget validation. CVMix, too, does not yet follow the working precision. Isolating precision effects requires a fixed domain decomposition, because a different decomposition alone changes the initial global-mean salinity by \saltDecompShift\,mpsu. The model-wide precision parameter also allows 16-bit formats to be tested in selected kernels \citep{klower2020formats, klower2022float16, ackmann2022solvers}, which would require the same budget diagnostics.

%% file: sections/07_conclusions.tex
\conclusions
\label{sec:conclusions}

FESOM2 can run forced climate integrations in single precision. Over
60 years at nominal $1^\circ$ resolution, its heat content, temperature, and
mean state differ from double precision about as much as two double-precision
runs with slightly perturbed initial conditions, its sea ice by at most
\finIceMaxPct\,\%, and the hindcasts run \finProdSpeedup\ times faster.
Coupled to an atmosphere, it adds no detectable heat source or sink over
\awicmRunYears\ years. On meshes from $1^\circ$ to 5\,km, the Fortran model runs up to \speedupFortranCpuMax\ times faster, and on
GPUs the C++/Kokkos version needs \gpuMemSavingMin--\gpuMemSavingMax\,\% less
memory, which may matter as much as speed at kilometre scale.

This required more than switching the working precision. A first
implementation lost salt, developed a heat-content deficit carried by the
surface fluxes, and shifted ocean fronts, all from rounding errors of order
$10^{-7}$ accumulated over decades. Storing salinity as an anomaly reduced the
drift by a factor of \saltAnomGain\ but did not remove it. Accumulating tracer
tendencies and layer thicknesses in 64~bits and carrying the rounding error
of each tracer update into the next did, at up to
\stackCostFortranCpuMaxPct\,\% additional cost; the remaining salinity
difference is \finSaltReduction\ times smaller and enters through the surface
flux.

Reduced precision is therefore a matter of numerical design and
verification, not a compiler switch. How prognostic variables are updated
decides whether small increments survive; conserved budgets reveal a
precision-induced source or sink within \emergeBudgetYear~model year, long
before its climatic consequences; and comparing independent implementations
exposes errors that double-precision tests miss.

Eddy-rich and kilometre-scale configurations, century-scale conservation,
biogeochemical tracers, and 16-bit kernels remain to be evaluated. These steps call for
precision-aware Earth-system models, in which working precision, variable
representation, and the placement of accumulations are designed together with
the numerical algorithms and the hardware.

%% file: sections/99_availability.tex
\codedataavailability{
FESOM2 is available at \url{https://github.com/FESOM/fesom2} under the GNU
General Public License v3.0, with releases archived on Zenodo. Single working
precision was introduced in FESOM~2.8.0 \citep{fesom2v280}. Pull request 940
was merged on 28 August 2026 (commit \expMergeCommit); subsequent changes
include the salinity anomaly (977, 986), solver defaults (984, 995, 997),
64-bit accumulation of output means (1032), tracer-conservation correction
(1054 in 2.8.0, extended to both precisions by 1056 in 2.8.1), and 64-bit
tracer tendencies and layer thicknesses with compensated tracer updates
(1206, 1207).

The Fortran source archive \citep{spcodefortran2026} contains mainline
revision \finSha\ for the final 60-year experiments, the component timings,
and the scaling measurements, and revision \expRunSha\ for the earlier six
60-year experiments. Revision \expRunSha\ (19 August 2026) was on the
integration branch of pull request 940 before a rebase and is preserved as tag
\texttt{sp-paper-60yr-27ed9f4f} at
\url{https://github.com/koldunovn/fesom2}; revision \finSha\ is tagged
\texttt{sp-paper-60yr-50467fa0} there. The remaining runs use neighbouring
revisions of the same code: mainline revision \oracleSha\ (included in 2.8.0)
for the implementation comparison and the Fortran scaling points at 256 and
512 nodes, and earlier revisions with the change under test for the controlled
experiments; \citet{spdata2026} maps each result to its source revision.

The C++/Kokkos implementation is available at
\url{https://github.com/koldunovn/fesom_kokkos} under the Apache License 2.0.
Revisions \portLadderSha, used in Sect.~\ref{sec:performance}, and
\portBinSha, used in Sect.~\ref{sec:port}, are archived with their Kokkos 4.4.01
dependency \citep{trott2022kokkos} in \citet{spcodekokkos2026}.

Run configurations and job scripts, the partitioning recipe, reduction and
plotting scripts, and the derived data used to generate all figures and
numerical results (global time series, climatologies, benchmark timings,
implementation-comparison statistics, and provenance records) are archived in
\citet{spdata2026} under the MIT licence.

Mesh archives are available for fArc \citep{wang2026farcmesh}, DARS
\citep{sein2026darsmesh}, and NG5 \citep{koldunov2025ng5mesh}. The NG5
vertical-level files used here differ from the archived version at one node
and two elements; our copies are included in \citet{spdata2026}. The CORE2
mesh, PHC3.0-derived initial state \citep{steele2001phc}, and 1958 JRA55-do
forcing used in Sect.~\ref{sec:port} are archived in
\citet{koldunov2026core2}. The 60-year experiments use CORE-II interannual
forcing \citep{large2009bulk,yeager2008core2data}; all other runs use
JRA55-do v1.4.0 \citep{tsujino2018jra55do,tsujino2019jra55do140}.
HadISST1 \citep{rayner2003hadisst} and WOA09 salinity
\citep{antonov2010woa09} were obtained from the ICDC data pool at
Universit\"at Hamburg.

The coupled experiments of Sect.~\ref{sec:coupled} use AWI-ESM3 version 3.5.1
with FESOM~2.8.3 and the halo exchange of the initial-condition perturbation
from FESOM pull request 1214, set up with ESM-Tools \citep{barbi2021esmtools}.
Their run configurations and reduction scripts are archived in
\citet{spdata2026}.

Raw output from the six 60-year experiments totals about 1.3\,TB and exceeds
the archive's capacity. It is retained on DKRZ Levante, is available from the
authors on request, and can be regenerated from the archived code,
configurations, and input data.
}

%% file: sections/A1_appendix.tex
\section{Mesh characteristics}
\label{app:meshes}

Table~\ref{tab:meshes} lists the four meshes used for performance measurements,
with dimensions read from the model's mesh files. CORE2 is the nominal
$1^\circ$ mesh used in the climate experiments. fArc refines CORE2 to
4.5\,km in the Arctic \citep{wang2018arctic}. DARS follows the local Rossby radius
\citep{sein2017rossby} and resolves energetically active regions more finely.
NG5 has nearly uniform 5\,km resolution over most of the globe. The larger
meshes also have more vertical levels, so their computational size grows
faster than their surface-vertex count.

\begin{table}[h]
\caption{Meshes used in the performance measurements. Vertices and triangles are
surface counts; levels are vertical grid levels. Resolution describes the mesh
design rather than an area-weighted mean.\label{tab:meshes}}
\begin{tabular}{lrrrl}
\tophline
Mesh & Vertices & Triangles & Levels & Resolution \\
\middlehline
CORE2 & 126\,858   & 244\,659    & 48 & nominal $1^\circ$ \\
fArc  & 638\,387   & 1\,253\,306 & 48 & CORE2 + 4.5\,km Arctic \\
DARS  & 3\,160\,340 & 6\,262\,485 & 57 & Rossby-radius scaled \\
NG5   & 7\,402\,886 & 14\,741\,520 & 70 & $\sim$5\,km quasi-uniform \\
\bottomhline
\end{tabular}
\belowtable{}
\end{table}

\section{Scaling configurations}
\label{app:campaign}

The measurements run on DKRZ Levante, whose CPU nodes have two 64-core AMD
EPYC 7763 processors (128 processes per node) and whose GPU nodes have four
NVIDIA A100 devices (one process per device). The Fortran model is compiled
with the Intel compiler. All runs use the physical settings of the Kokkos
implementation's default CORE2 configuration on every mesh: linear free
surface, biharmonic horizontal viscosity, KPP mixing, standard EVP sea ice,
split vertical velocity on fArc, DARS, and NG5, and symmetric
sea-surface-height preconditioning in both precisions. They use JRA55-do
forcing, PHC3.0 initial conditions, and timesteps of 1800\,s for CORE2,
900\,s for fArc, 120\,s for DARS, and 180\,s for NG5.

Each job runs double precision, single precision, and single precision
without conservative updates twice, in mirrored order; Kokkos jobs start with
a discarded warm-up run. Each run starts from the initial state, advances
300 timesteps without output, and is timed by the implementation's own step
timer; we report the faster of the two runs of each build. Repeated runs
differ by a median of \legSpreadMedianPct\,\% and at most
\legSpreadMaxPctRest\,\%, except for DARS at the two largest process counts,
where they differ by up to \legSpreadMaxPct\,\%; the apparent increase of the
DARS speedup between 256 and 512 nodes lies within this spread, and the step
time at 512 nodes shows that DARS has passed its useful scaling range there.
Speedup is the double-precision time divided by the single-precision time for
the same code, mesh, and process count. Absolute times of the two
implementations are indicative, because their compilers and timers differ.
Throughput is projected to production timesteps of 1800\,s for CORE2, 1200\,s
for fArc, and 240\,s for DARS and NG5, neglecting a 1--2\,\% increase in cost
per step at the longer timestep that is measured in the Kokkos implementation
and caused by additional solver iterations.

Fortran partitions are generated at each process count with the model's
METIS partitioner \citep{karypis1998metis} and two-dimensional vertex
weights, following the recipe supplied with the meshes; on CORE2 at
864 processes, this reproduces the step time of the supplied partition within
\partRecipeMatchPct\,\%. The Kokkos runs use the supplied partitions.

CPU process counts range from 64--4096 for CORE2, 512--8192 for fArc,
1024--65\,536 for DARS, and 2048--65\,536 for NG5, and GPU counts from 4--64,
8--64, 8--128, and 16--128. The lower limit is set by memory; CPU runs extend
to 512 nodes, the allocation limit of the partition. The DARS and NG5 Fortran
points at 256 and 512 nodes come from an earlier campaign with a version
without conservative updates, which used the same protocol without the third
build; two of them, DARS at 32\,768 and NG5 at 65\,536 processes, use the
completed runs of jobs that reached their time limit. At process counts
measured in both campaigns, absolute double-precision times differ by
$\ladderCampaignDpDiffMinPct$ to $+\ladderCampaignDpDiffMaxPct$\,\%, most at
the highest counts, so the absolute times of the earlier points are
indicative, whereas their speedups are ratios within one job. In the DARS job
at 16\,384 processes, one single-precision run failed at start-up on a
network error, and the time of that build is the remaining run.

%% file: sections/A2_before.tex
\section{Single precision before the conservative updates}
\label{app:before}

This appendix documents the 60-year hindcasts with the first implementation
(Sect.~\ref{sec:exp-pair}), which precede the conservative updates of
Sect.~\ref{sec:conservation}, and the controlled experiments that identified
the source of their salinity drift. Section~\ref{sec:before} summarizes them.
The double-precision control of these hindcasts is a pair of runs that differ
only in the salinity representation, and hence only in round-off.

\subsection{Global budgets and sea ice}
\label{sec:timeseries}
\label{app:before-budgets}

\begin{figure*}[tp]
\centering
\includegraphics[width=0.86\textwidth]{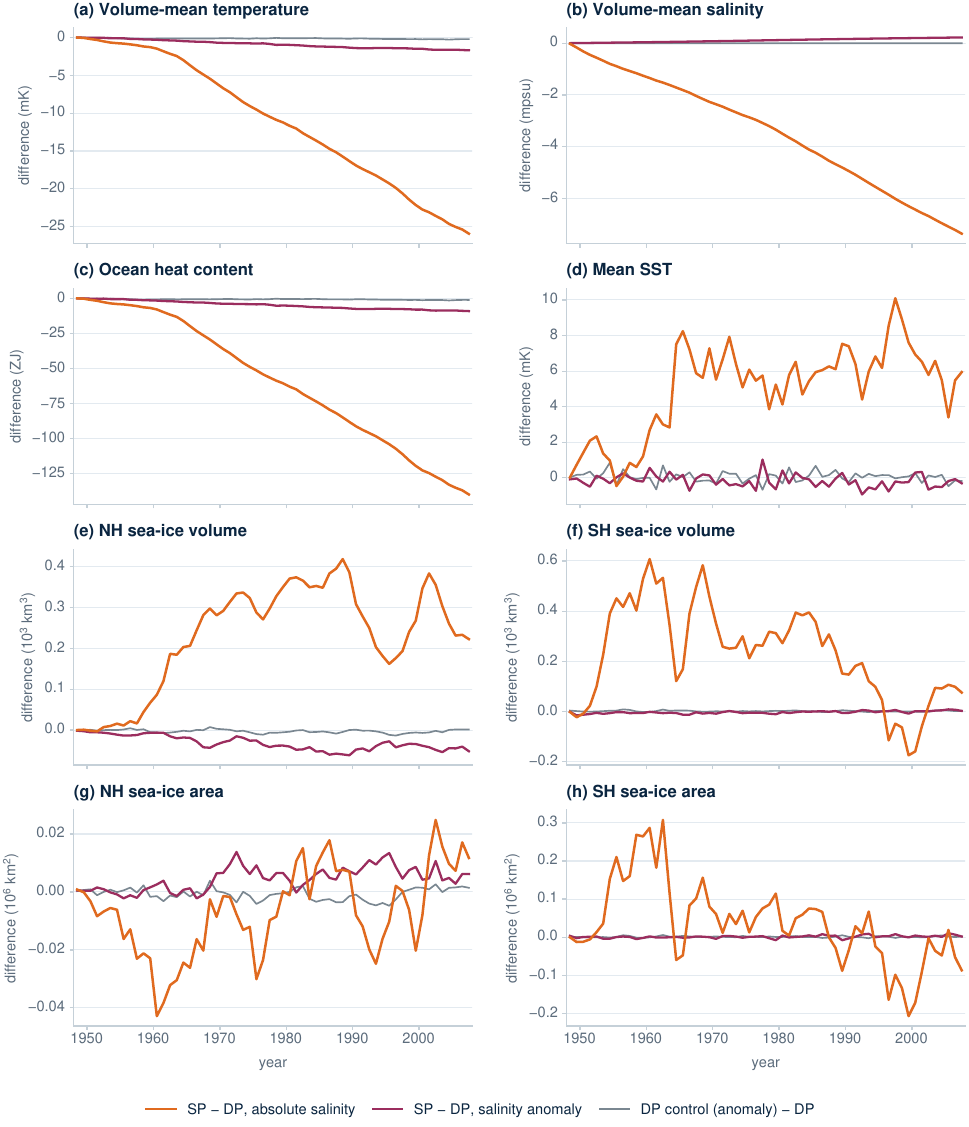}
\caption{Annual global-integral differences in the 60-year CORE2 hindcasts
with the first implementation, without the conservative updates of Sect.~\ref{sec:conservation}.
Orange and magenta show single minus double precision with absolute salinity
and with the anomaly, respectively. Gray shows the difference between the
double-precision salinity representations.
\label{fig:timeseries}}
\end{figure*}

Figure~\ref{fig:timeseries} shows the annual global integrals of the first
implementation as differences between paired runs. With absolute salinity,
single precision loses salt at a steady $\saltLossPerYearSci$\,psu\,yr$^{-1}$
and reaches a last-decade deficit of \saltDriftSpDp\,mpsu, whereas the control
differs by about $10^{-4}$\,mpsu and the double-precision run itself changes by
\dpOwnSaltTrend\,mpsu over sixty years; an earlier pair with pre-merge code on
128 processes gave a similar deficit of \saltDriftOldPair\,mpsu. The ocean also
loses \ohcGapSpDp\,ZJ of heat, and its volume-mean temperature falls by
\tempDriftSpDp\,mK, about a third of the double-precision spin-up warming of
\dpOwnTempTrend\,mK, although SST is slightly higher.

With the salinity anomaly, the salinity trend reverses sign and shrinks by a
factor of \saltAnomGain, to $\saltTrendSpDpAnomPerYear\times10^{-6}$\,psu\,yr$^{-1}$,
and the heat-content and temperature deficits shrink to \ohcGapSpDpAnom\,ZJ
(a reduction of \ohcAnomRemovedPct\,\%) and \tempDriftSpDpAnom\,mK. These
residuals are still systematic: the salinity difference exceeds that of the
control by more than three orders of magnitude, and the temperature and
heat-content differences are several times the control.

The heat budget separates the heat-content differences into a surface-flux
response and an internal residual. Integrating the monthly surface heat flux
and the heat carried by the freshwater flux closes the budget of each
double-precision run to within \dpBudgetClosureWm\,W\,m$^{-2}$. With absolute
salinity, \ohcSurfaceSpDp\,ZJ of the deficit comes from the surface fluxes and
\ohcInternalSpDp\,ZJ from the internal residual. With the anomaly, the surface
fluxes supply \ohcSurfaceSpDpAnom\,ZJ more heat than in double precision, while
an internal sink removes \ohcInternalSpDpAnom\,ZJ, or
\heatSinkSpAnomWm\,W\,m$^{-2}$; the internal residual of the control stays
within \ohcInternalFloorMax\,ZJ. The salinity differences and the internal
sinks exceed the control by model year~\emergeBudgetYear,
whereas the heat-content difference of the anomaly configuration does so only
from year~\emergeOhcAnomYear. Section~\ref{sec:mechanism} attributes most of
the internal sink with the anomaly to the rounding of temperature updates
rather than to the FCT defect described in Sect.~\ref{sec:conservation}.

With absolute salinity, Arctic ice volume is \IceVolNhSpDpPct\,\% larger
in the last decade, with a maximum difference of \IceVolNhSpDpMaxPct\,\%,
while ice area is essentially unchanged. Antarctic ice volume and area show
a transient response that peaks around 1960, reaching
\IceVolShSpDpMaxPct\,\% in volume, and subsides by the end of the record.
With the anomaly, volume differences remain within \IceVolNhAnomMaxPct\,\%
in the Arctic and \IceVolShAnomMaxPct\,\% in the Antarctic; the corresponding
baseline maxima are \IceVolNhFloorMaxPct\,\% and
\IceVolShFloorMaxPct\,\%.

\subsection{Source of the salinity drift}
\label{sec:mechanism}

\begin{figure}[tp]
\includegraphics[width=8.3cm]{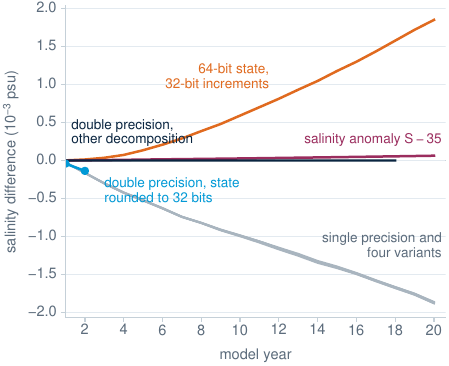}
\caption{Global-mean salinity differences in controlled experiments with
pre-merge code at 512 processes, relative to an earlier double-precision run
at 128 processes. Gray: single precision and variants with a different
decomposition, 64-bit global sums, 64-bit solver operations, or small state
perturbations. Orange: 64-bit tracer state with 32-bit increments. Cyan:
double precision with the tracer state rounded to 32~bits after each step
(two years). Magenta: salinity stored as $S-35$. Dark blue: double precision
with a different decomposition. \label{fig:twindrift}}
\end{figure}

The surface freshwater budget does not explain the freshening. Monthly net
fluxes close to within each build's machine precision, with no systematic
bias or trend, and are \fluxMarginOrders\ orders of magnitude smaller than
the flux required to explain the drift. Controlled 20-year experiments also
exclude global budget reductions, the solver, and domain decomposition as
its main sources (Fig.~\ref{fig:twindrift}). Using 64-bit budget reductions
or solver operations, changing the decomposition, or adding perturbations
smaller than the spacing between representable salinity values yields nearly
the same drift as the unmodified single-precision run:
$\twinVanillaTwenty$\,mpsu at year 20. A double-precision control on a different
decomposition remains within \twinDpCtlMax\,mpsu of the reference.

The experiments instead implicate rounding in the salinity update,
$S\leftarrow S+\Delta S$. At $S\approx35$, increments smaller than
$1.9\times10^{-6}$\,psu can be lost when added to a 32-bit state, and the
calculation of the increments can itself introduce bias. This loss of small
increments, known as stagnation \citep{croci2023heat}, has biased slowly
evolving deep-soil temperatures in single-precision land-surface models
\citep{harvey2016soil,dawson2018land}, and single-precision tracer advection
in the MRI.COM ocean model lost small tracer fluxes added to the tracer content of grid
cells, producing temperature errors that grew over 300 years
\citep{nakano2025som}. Retaining the state
in 64~bits while calculating increments in 32~bits produces a drift of
$+\twinLedgerTwenty$\,mpsu after 20 years: nearly the same magnitude as the
unmodified run, but of opposite sign. Conversely, rounding the tracer state
to 32~bits after each double-precision timestep produces a drift of
$\twinQuantTwo$\,mpsu in two years. Thus, increasing storage precision alone
does not eliminate the bias. This differs from the land-surface,
shallow-water, and ocean-advection cases above, in which a higher-precision
state largely removes the error
\citep{dawson2018land,klower2020formats,nakano2025som}: here the increments are themselves
calculated from salinities near 35\,psu, and their rounding is biased. The
anomaly representation reduces rounding in both the updates and the
increments and gives a 20-year difference of \twinAnomTwenty\,mpsu,
\twinAnomSuppression\ times smaller in magnitude than the unmodified run,
with little computational overhead (Sect.~\ref{sec:anatomy}).

Because the salinity differences and internal heat sinks emerge within the
first model year (Sect.~\ref{sec:timeseries}), nine-month restarts from the
hindcast states at the end of 1997 can separate their sources. Correcting the
FCT defect (Sect.~\ref{sec:conservation}) reduces the salt
loss with absolute salinity by \sxFctAbsReductionPct\,\% but changes the
residuals with the anomaly by about \sxFctAnomChangePct\,\%; FESOM~2.8.1,
which includes the correction, gives an internal heat sink of
\sxSinkWm\,W\,m$^{-2}$ and a salinity gain of $\sxSaltUpsu\times10^{-6}$\,psu
in nine months. Compensated (Kahan) summation \citep{kahan1965} of the
tracer updates alone turns these residuals into a source of
\sxKahanSourceWm\,W\,m$^{-2}$ and a loss of
$\sxKahanSaltLossUpsu\times10^{-6}$\,psu, both still above the difference
between the double-precision restarts; with absolute salinity, it leaves a
gain of $\sxAbsKahanSaltUpsu\times10^{-6}$\,psu, as 64-bit storage does.
These results motivated computing the increments in 64~bits as well and
storing layer thicknesses at that precision (Sect.~\ref{sec:conservation}).

\subsection{Mean-state differences and observational biases}
\label{sec:yardstick}

\begin{figure*}[tp]
\includegraphics[width=\textwidth]{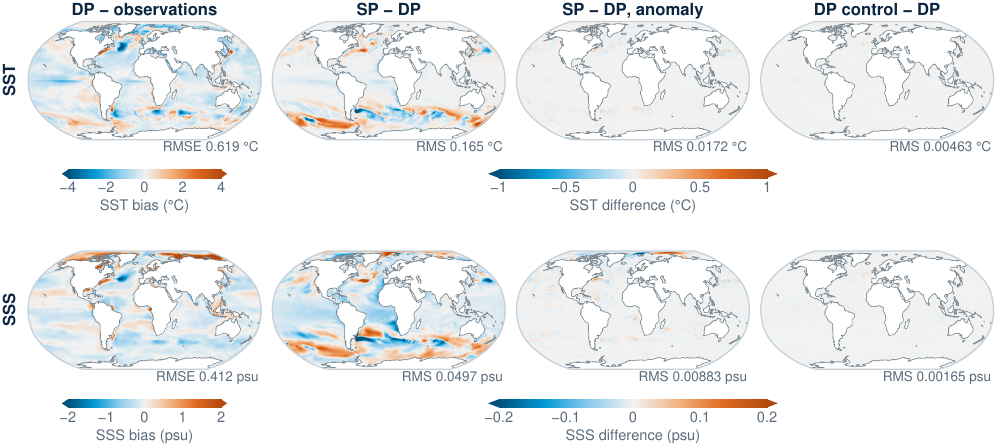}
\caption{Mean-state differences for 1988--2007 with the first implementation. Left: double-precision bias
against observations (SST against HadISST1 and surface salinity against WOA09),
with area-weighted root-mean-square error. Remaining columns: single minus
double precision with absolute salinity, the same with the salinity anomaly,
and the difference between the double-precision salinity representations.
These three columns use a common color scale and show area-weighted RMS
differences.
\label{fig:meanstate}}
\end{figure*}

We compare the 1988--2007 mean states with both the double-precision baseline
and the model's bias against independent observational products
(Fig.~\ref{fig:meanstate}): HadISST1 for SST \citep{rayner2003hadisst} and the
World Ocean Atlas 2009 (WOA09) for surface salinity \citep{antonov2010woa09}.
PHC3.0 supplies the initial conditions and is not used for this evaluation.

With absolute salinity, the root-mean-square (RMS) single-minus-double
precision differences are \yardSstRmsDiff\,$^\circ$C for SST and
\yardSssRmsDiff\,psu for surface salinity. These are \yardSstRatio\ and
\yardSssRatio\ of the respective model errors against observations, but
\yardSstOverFloor\ and \yardSssOverFloor\ times the double-precision
baseline. The spatial pattern follows major fronts in the North Atlantic
Current, Southern Ocean, and Brazil--Malvinas Confluence, indicating a
systematic mean-state shift.

With the anomaly, the RMS differences decrease to
\yardSstRmsDiffAnom\,$^\circ$C and \yardSssRmsDiffAnom\,psu, or
\yardSstRatioAnom\ and \yardSssRatioAnom\ of the errors against
observations. The coherent frontal pattern disappears. The remaining
differences are \yardSstOverFloorAnom\ and \yardSssOverFloorAnom\ times
the baseline: small relative to observational biases, but still distinguishable
from the double-precision control.

\begin{figure*}[tp]
\includegraphics[width=\textwidth]{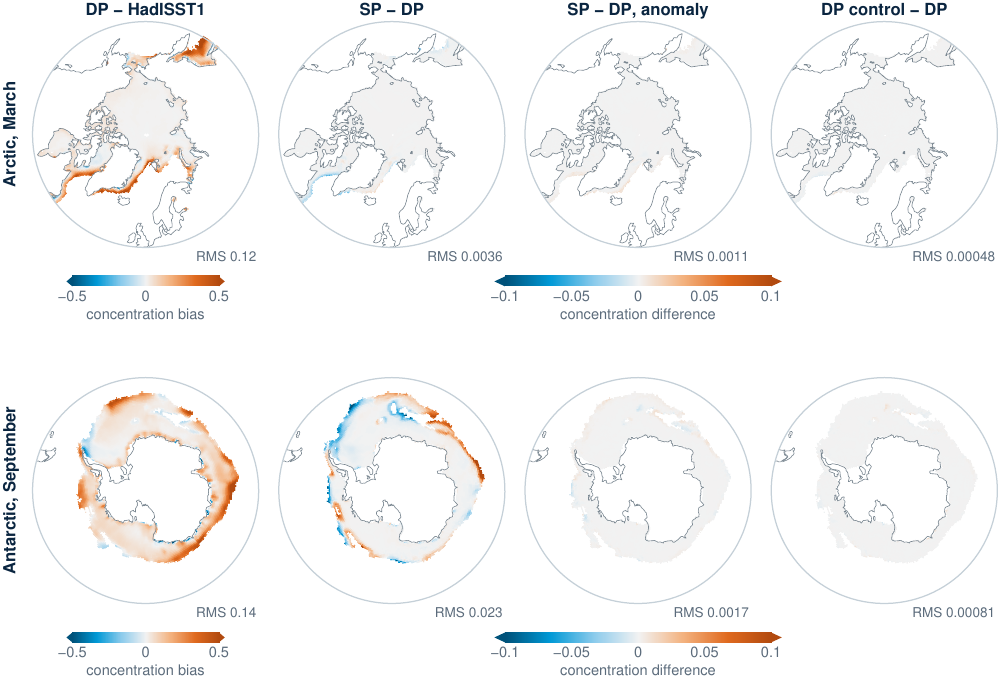}
\caption{Mean sea-ice concentration differences for 1988--2007 at the seasonal
maxima (March in the Northern Hemisphere and September in the Southern
Hemisphere). Columns follow Fig.~\ref{fig:meanstate}. RMS differences are
calculated over nodes where modeled or observed concentration exceeds 0.05.
\label{fig:seaice}}
\end{figure*}

Sea-ice concentration shows a similar response (Fig.~\ref{fig:seaice}). With
absolute salinity, precision-induced differences are
\yardIceRatioMin--\yardIceRatioMax\ of the model error against HadISST1 and
up to \yardIceOverFloorMax\ times the baseline, concentrated near the
Antarctic ice edge. With the anomaly, they fall to at most
\yardIceRatioAnomMax\ of the model error against observations and
\yardIceOverFloorAnomMax\ times the baseline. Ice thickness improves less:
the residual Arctic volume difference is about a fifth of a percent and
remains several times the baseline (Fig.~\ref{fig:timeseries}).

%% file: sections/A3_coupled.tex
\section{Coupled integrations}
\label{app:coupled}

\begin{table}[!htbp]
\caption{Coupled AWI-ESM3 integrations: differences between single and double
precision and between the perturbed runs and their references. The internal
heat residual is the 60-year mean of the part of the heat-content change not
supplied by the surface fluxes. All other entries are last-decade (1900--1909)
annual means.\label{tab:coupled}}
\begin{tabular}{lccc}
\tophline
 & SP $-$ DP & DP twin $-$ DP & SP twin $-$ SP \\
\middlehline
$|\Delta|$ internal heat residual (W\,m$^{-2}$) & $\awicmInternalSpDp$ & $\awicmInternalTwinDp$ & $\awicmInternalTwinSp$ \\
Heat content (ZJ) & $\awicmOhcSpDp$ & $\awicmOhcTwinDp$ & $\awicmOhcTwinSp$ \\
Global-mean SST (K) & $\awicmSstSpDp$ & $\awicmSstTwinDp$ & $\awicmSstTwinSp$ \\
Global-mean 2\,m temperature (K) & $\awicmTasSpDp$ & $\awicmTasTwinDp$ & $\awicmTasTwinSp$ \\
Net radiation at the top (W\,m$^{-2}$) & $\awicmToaSpDp$ & $\awicmToaTwinDp$ & $\awicmToaTwinSp$ \\
Arctic sea-ice volume (\%) & $\awicmIceNhSpDp$ & $\awicmIceNhTwinDp$ & $\awicmIceNhTwinSp$ \\
Antarctic sea-ice volume (\%) & $\awicmIceShSpDp$ & $\awicmIceShTwinDp$ & $\awicmIceShTwinSp$ \\
\bottomhline
\end{tabular}
\belowtable{}
\end{table}

The coupled experiments of Sect.~\ref{sec:coupled} use AWI-ESM3 version 3.5.1,
the successor of AWI-CM3 \citep{streffing2022awi}. FESOM~2.8.3
runs on the CORE2 mesh of Sect.~\ref{sec:model} with the same vertical
coordinate, mixing, and eddy parameterizations, but with a 1800\,s timestep,
without surface salinity restoring, and without icebergs. The atmosphere model
OpenIFS 48r1 \citep{ecmwf2024openifs} runs at TCo95 resolution with 91 levels
and a 1\,h timestep, and the dynamic vegetation model LPJ-GUESS
\citep{smith2014lpjguess} on the land points of the OpenIFS grid. The coupler
OASIS3-MCT \citep{craig2017oasis} exchanges fluxes and surface state between
atmosphere and ocean every 2\,h and between atmosphere and vegetation once a
day. The fields that FESOM2 passes to OASIS are 64-bit in both precisions.

All four runs use 1850 forcing and start cold: the ocean at rest from PHC3.0,
the atmosphere from its standard initial state, and LPJ-GUESS without a
spun-up state. FESOM2 runs on 512 processes, OpenIFS on 192 processes with
four threads each, and LPJ-GUESS on 64 processes, 13 nodes in total, with the
same layout and coupling weights in all four runs. The two perturbed runs add
one Gaussian draw with a standard deviation of $2\times10^{-4}$\,K to the
initial temperature of each water column, with the same random seed in both
precisions. FESOM2 now exchanges the halos of the perturbed fields. Without
this exchange, the perturbed runs did not conserve volume, heat, or salt. The model
records its budgets every simulated day, as in Sect.~\ref{sec:exp-pair}. In
double precision, the volume and salt residuals stay below
$\awicmResidDpMax$ of the content. In single precision, a restart changes the
heat content by $\awicmHeatJump$ of its value, as in the forced runs.

%% file: sections/99b_statements.tex
\authorcontribution{
SKC implemented single working precision in FESOM2 and ran the 60-year
experiments. NK designed the study, implemented single precision in the
C++/Kokkos version, and carried out the analysis and the performance
measurements. JS ran the coupled experiments. SD and DS contributed to the
conceptualization. TJ acquired funding and managed the project. All authors
contributed to writing the manuscript.
}

\competinginterests{
The authors declare that they have no conflict of interest.
}

\begin{acknowledgements}
We thank the German Climate Computing Center (DKRZ) for access to its computing
infrastructure.
\end{acknowledgements}

\financialsupport{
This work was primarily supported through the core funding of the Alfred Wegener
Institute, Helmholtz Centre for Polar and Marine Research (AWI), within the Helmholtz
Association. SKC is funded by the WarmWorld Better project of the German Federal
Ministry of Research, Technology and Space under the funding code 01LK2202A; the
responsibility for the content of this publication lies with the authors. The
contribution by NK was supported by the TerraDT (Digital Twin of Earth system for
Cryosphere, Land surface and related interactions) project, which has received funding
from the European Union's Horizon Europe research and innovation programme under Grant
Agreement no.~101187992. SD was supported by project M5 of the
Collaborative Research Centre TRR181 ``Energy Transfer in Atmosphere and Ocean'' funded
by the Deutsche Forschungsgemeinschaft (DFG, German Research Foundation) --
Projektnummer 274762653. TJ and NK were supported by the EERIE project (Grant Agreement
No~101081383) funded by the European Union. Views and opinions expressed are however
those of the author(s) only and do not necessarily reflect those of the European Union
or the European Climate Infrastructure and Environment Executive Agency (CINEA).
Neither the European Union nor the granting authority can be held responsible for them.
The work of SKC and NK has been supported by the European Union's Destination Earth
Initiative and relates to tasks entrusted by the European Union to the European Centre
for Medium-Range Weather Forecasts implementing part of this Initiative with funding by
the European Union. Views and opinions expressed are those of the authors only and do
not necessarily reflect those of the European Union or the European Commission. Neither
the European Union nor the European Commission can be held responsible for them.
}